\documentclass[showpacs,preprintnumbers,amsmath,amssymb,APSl,prd,nofootinbib,superscriptaddress]{revtex4-1}

\usepackage{graphicx,color}
\usepackage{subfig}
\usepackage{amssymb}
\usepackage{hyperref}
\usepackage[utf8]{inputenc}
\usepackage[english]{babel}
\usepackage{epsfig}
\usepackage{wasysym}
\usepackage{color,xcolor}
\usepackage{amsmath}
\usepackage{bm}
\usepackage{epsfig}
\usepackage{amsfonts}
\usepackage{dcolumn}
\usepackage{float}

\begin{document}

\title{ Gravitational lens effect of a holonomy corrected Schwarzschild black hole\\}


	\author{Ednaldo L. B. Junior} \email{ednaldobarrosjr@gmail.com}
\affiliation{Faculdade de Engenharia da Computação, Universidade Federal do Pará, Campus Universitário de Tucuruí, CEP: 68464-000, Tucuruí, Pará, Brazil}

	\author{Francisco S. N. Lobo} \email{fslobo@ciencias.ulisboa.pt}
\affiliation{Instituto de Astrof\'{i}sica e Ci\^{e}ncias do Espa\c{c}o, Faculdade de Ci\^{e}ncias da Universidade de Lisboa, Edifício C8, Campo Grande, P-1749-016 Lisbon, Portugal}
\affiliation{Departamento de F\'{i}sica, Faculdade de Ci\^{e}ncias da Universidade de Lisboa, Edif\'{i}cio C8, Campo Grande, P-1749-016 Lisbon, Portugal}

	\author{Manuel E. Rodrigues} \email{esialg@gmail.com}
\affiliation{Faculdade de F\'{i}sica, Programa de P\'{o}s-Gradua\c{c}\~{a}o em F\'{i}sica, Universidade Federal do Par\'{a}, 66075-110, Bel\'{e}m, Par\'{a}, Brazill}
\affiliation{Faculdade de Ci\^{e}ncias Exatas e Tecnologia, Universidade Federal do Par\'{a}, Campus Universit\'{a}rio de Abaetetuba, 68440-000, Abaetetuba, Par\'{a}, Brazil}

	\author{Henrique A. Vieira} \email{henriquefisica2017@gmail.com}
\affiliation{Faculdade de F\'{i}sica, Programa de P\'{o}s-Gradua\c{c}\~{a}o em F\'{i}sica, Universidade Federal do Par\'{a}, 66075-110, Bel\'{e}m, Par\'{a}, Brazill}


\begin{abstract}

In this paper we study the gravitational lensing effect for the Schwarzschild solution with holonomy corrections. We use two types of approximation methods to calculate the deflection angle, namely the weak and strong field limits. For the first method, we calculate the deflection angle up to the fifth order of approximation and show the influence of the parameter $\lambda$ (in terms of loop quantum gravity) on it. In addition, we construct expressions for the magnification, the position of the lensed images and the time delay  as functions of the coefficients from the deflection angle expansion. We find that $\lambda$ increases the  deflection angle. 
In the strong field limit, we use a logarithmic approximation to compute the deflection angle. 
We then write four observables, in terms of the coefficients $b_1$, $b_2$ and $u_m$, namely: the asymptotic position approached by a set of images $\theta_{\infty}$, the distance between the first image and the others $s$, the ratio between the flux of the first image and the flux of all other images $r_m$, and the time delay between two photons $\Delta T_{2,1}$. We then use the experimental data of the black hole Sagittarius $A^{\star}$ and calculate the observables and the coefficients of the logarithmic expansion.  We find that the parameter $\lambda$ increases the deflection angle, the separation between the lensed images and the delay time between them. In contrast, it decreases the brightness of the first image compared to the others.

\end{abstract}

\date{\today}

\maketitle

\section{Introduction}

Shortly after Einstein published the field equations of general relativity \cite{Einstein1905}, Karl Schwarzschild \cite{SZ} proposed an exact solution to these equations, which became known as the Schwarzschild black hole \cite{Herdeiro:2018ldf}. The current definition of these bodies is a region of spacetime covered by an event horizon from which not even light can escape. 
Initially, black holes were discredited and many argued that they were just a mathematical solution with no relation to reality. However, they gained notoriety in the 1960s with the discovery of compact objects and more recently with the first image of the shadow of what is believed to be a supermassive black hole \cite{fotoBN1,fotoBN2,fotoBN3,fotoBN4,fotoBN5,fotoBN6}.
Over the years, many other exact solutions have appeared, such as those of Reissner-Nordstr\"o{}m and Kerr \cite{Herlt}, but like the Schwarzschild geometry, they suffered from a peculiarity that troubled the scientific community, i.e., the existence of a singularity. A curvature singularity is a sudden endpoint in the geodesic equations at which quantities such as the density of matter become infinite. In fact, the notion of geodesic completeness essentially resides at the root of the singularity theorems \cite{Hawking:1973uf} and is considered as a key element to determine the presence of a spacetime singularity \cite{Olmo:2017fbc}.

The main focus of physicists attempting to eliminate the singularity is the development of a theory of quantum gravity. Initial attempts date back to the middle of the last century \cite{DeWitt}, but we still have neither a satisfactory theory nor experimental data on the quantum aspects of gravity \cite{Ashtekar:2021kfp}. For instance, we refer the reader to \cite{Thiemann:2001gmi,Percacci:2023rbo} and references therein to get a compilation of modern quantum gravity, including its challenges and advances.
Alternatively to searching for a complete quantum description of gravitation there is the possibility to describe some phenomena at low energy scales (compared to the Plack scale) \cite{Donoghue:1994dn,Burgess:2003jk,Buchbinder:1992rb}.
Through these models it is also possible to find corrections from gravitation to quantum electrodynamics \cite{Bevilaqua:2021uev,Bevilaqua:2021uzk,Souza:2022ovu} and  to quantum chromodynamics (QCD) \cite{Souza:2023wzv}. Another interesting point  is the influence of quantum gravitation on dark matter \cite{Calmet:2009uz,Calmet:2021iid}. 
Among the various effective theories, the so-called Loop Quantum Gravity (LQG) has gained notoriety.
In cosmology, LQG has been used in the creation of cosmological models without singularity
\cite{Bojowald:2001xe,Saini:2018tto,Saini:2017ggt}, explanation of the Big Bang \cite{Ashtekar:2006wn,Varadarajan:2008bh}, and other effects \cite{Bojowald:2005epg,Ashtekar:2011ni}. 
In black hole physics,  Ashtekar and  collaborators \cite{Ashtekar:1997yu} used the LQG formalism and obtained a new derivation for the  Bekenstein-Hawking formula, Vakili \cite{Vakili:2018xws} use the Schwarzschild metric as background geometry in the framework of classical polymerization and showed that its energy-momentum tensor has the features of dark energy.
Several other papers have also studied LQG corrections to the Schwarzschild solution \cite{Boehmer:2007ket,Chiou:2008nm,BenAchour:2018khr,Bodendorfer:2019cyv}.
 Recently, Alonso-Bardaji, Brizuela and Vera \cite{Alonso-Bardaji:2021yls,Alonso-Bardaji:2022ear} used a canonical transformation and a linear combination of the general gelativity constraints to propose a black hole model consisting of an LQG correction to the Schwarzschild solution. It is an anomaly-free model described by the following metric
\begin{equation}
    ds^2 = - A(r) dt^2 + \Bigl[ \left(1 - \frac{l}{r} \right)A(r) \Bigr]^{-1}dr^2 + r^2 \left(d\theta^2 + \sin^2 \theta d\phi^2  \right),
    \label{Schwholonomy}
\end{equation}
where $A(r)=1-2m/r$ is the usual Schwarzschild metric function, and $l$ is a new scale length defined by $l: = 2m\lambda^2/(1+\lambda^2)$,
where $\lambda$ is called the polymerization constant and provides the holonomy correction information. 
For $m > 0$, this solution is asymptotically flat and contains a globally hyperbolic black hole or white hole region with a minimal space-like hypersurface replacing the original singularity.
The full information on the event horizons, Penrose diagram, and spacetime structure associated with this model can be found in the original papers mentioned earlier.
In addition, it was show in \cite{Moreira:2023cxy} (a work focused on  quasi-normal modes) that  perturbations become less damped as we increase the LQG parameter $\lambda$.

The purpose of this work is to study the light deflection properties of the solution present in \cite{Alonso-Bardaji:2021yls,Alonso-Bardaji:2022ear}. The deflection of light rays by a massive body can produce a widely known effect, namely, gravitational lensing. At first, physicists, including Einstein, believed that this effect could be observed only in experiments such as the one in 1919 \cite{Sobral}. At that time, astronomers could  measure the deflection angle caused by the Sun during a solar eclipse \cite{Crispino}. In fact, this kind of observation is possible only with the Sun. This reasoning began to change with the pioneering ideas of Fritz Zwicky, who proposed that we can observe lensing effects caused by galaxies and even clusters of galaxies. For a more detailed overview of the early measurements and theoretical proposals on this topic, see, for example, \cite{Congdon,Kayser,Virbhadra:1999nm,Virbhadra:2002ju,Claudel:2000yi} and the references therein. 
Recent developments related to this effect can be divided into two groups: strong lensing and microlensing \cite{Congdon}.
Strong lensing is related to measurements of galaxies and clusters of galaxies and has gained notoriety since the discovery of the accelerating expansion of the Universe \cite{SupernovaSearchTeam:1998fmf,SupernovaCosmologyProject:1998vns}. The reason is that we do not know why the Universe is expanding, and a popular approach to explain this behavior is the presence of an exotic cosmic fluid denoted as dark energy \cite{Copeland:2006wr}. Although there is no direct measurement of dark energy, we have tempting evidence for its existence \cite{DES:2017gwu}, where the weak gravitational lensing effect plays an important role in these measurements \cite{Frieman:2008sn}. Much work has been done in the literature, for instance, Bartelmann and Schneider \cite{Bartelmann:1999yn} have shown how to obtain the deflection angle within these limits; Holz and Wald \cite{Holz:1997ic} presented a similar method for inhomogeneous universes; Lewis and Challinor \cite{Lewis:2006fu} commented on how the weak lensing effects affect the cosmic microwave background; and Ghaffarnejad and Niad \cite{Ghaffarnejad:2014zva} calculated this effect considering a Bardeen black hole. For an overview of these and other methods using weak lensing, we refer the reader to \cite{Weinberg:2013agg}.

Another motivation arising from strong lensing is the possibility of testing general relativity in a strong gravitational field (so far it has been tested only for weak fields). Bozza \cite{Bozza:2002zj} proposed a way to calculate the deflection angle by a logarithmic expansion. Later, Bozza and Mancini applied this formalist to Sgr A* to describe how to observe real black holes with GRAVITY \cite{Bozza:2012by}. Pietroni and Bozza, also considering Sgr A*, commented on the effect of gravitational lensing on stellar orbit reconstruction \cite{Pietroni:2022cur}. Naoki Tsukamoto reproduced Bozza's formalism considering a slightly unstable photon sphere \cite{Tsukamoto:2020iez}, then applied it to Simpson-Visser spacetime \cite{Tsukamoto:2020bjm} and to a Reissner-Nordstr\"om naked singularity \cite{Tsukamoto:2021fsz}. J. Zhang and Y. Xie later considered a black-bounce- Reissner-Nordstr\ "om solution \cite{Zhang:2022nnj}. Advances in this field is not limited to the theoretical realm, but there are also recent attempts to observe this effect for supermassive black holes \cite{Nightingale:2023ini,Legin:2022ovl}.
Microlensing is the measurement of the collective magnification of various images; in the case of stars, the observation may take months or even years \cite{Congdon}. This type of measurement can be used in the detection of bodies that do not emit light, such as planets \cite{Wambsganss:1996he,Bozza:2018loy} or black holes \cite{Sajadian:2023xsf}; and also helps in the study of binary star systems \cite{Shin:2012xz,Choi:2013ajr}. In addition, there are several other ways to apply gravitational lensing that are still in the theoretical realm. In \cite{Bozza:2003cp} the authors claim that it can be used as a distance estimator, and in \cite{Keeton:2005jd,Keeton:2006sa,Keeton:2006di} a formalism for testing gravitational theories using lenses through compact objects was proposed. In \cite{Horvath:2011xr}, gravitational lensing in the Kehagias-Sfetsos space-time \cite{Kehagias:2009is}, emerging in the framework of Hořava-Lifshitz gravity, has also been analysed.
In \cite{Poddar:2021sbc}, light bending serves as constraints on axionic fuzzy dark matter.

As mentioned above, in this work we will study the gravitational lensing effect for the Schwarzschild solution with holonomy corrections. We will consider two types of approximations, namely strong and weak lensing.
This paper is organized as follows: In Sec. \ref{sec:one}, we will briefly discuss the main aspects of the weak lensing system and then apply it to the above solutions. In Sec. \ref{sec:GB}; we will calculate again the first term of the weak field expansion using the Gauss-Bonnet theorem, and in Sec. \ref{sec:two} we repeat the same procedure as in the previous section, now considering strong lensing. In Sec. \ref{sec:obser} we numerically calculated the observavaies using data from the black hole at the center of our Galaxy. In Sec. \ref{sec:six} we discussed an extension of the photon surface concept to the case of massive particles
In Sec. \ref{sec:conclusion}, we draw our conclusions. We will use the metric signature $(-,+,+,+)$ in this paper. Also, unless otherwise stated, we will use geometrized units with $G = c =1$.

\section{Weak gravitational lensing \label{sec:one}}

\subsection{Setting the stage}

In this section, we compute the deflection angle and the observable (image position and magnification) for the Schwarzschild solution with holonomy corrections, given by Eq. (\ref{Schwholonomy}), in the weak field limit. In this regime, we assume that both the source and the observer are very far from the lens and the light rays are only slightly distorted by the lens.
Formally, we can define the weak field limit as follows:
\begin{itemize}
    \item The gravitational lens is compact, static, and spherically symmetric, with an asymptotically flat
 spacetime geometry far away from the lens. The spacetime is vacuum outside the lens and flat
 in the absence of the lens.
      \item The observer and the source lie in the asymptotically
flat regime of the spacetime.
        \item The distance scale is much larger than the mass scale, i.e. 
\begin{equation}
    \frac{m}{r_0} \ll 1, \ \ \ \ \frac{m}{u} \ll 1,
\end{equation}
where $r_0$ is the distance of closest approach to the lens and $u$ is the impact parameter. 
\end{itemize}

\begin{figure}[htpb]
\centering
\includegraphics[scale=0.5]{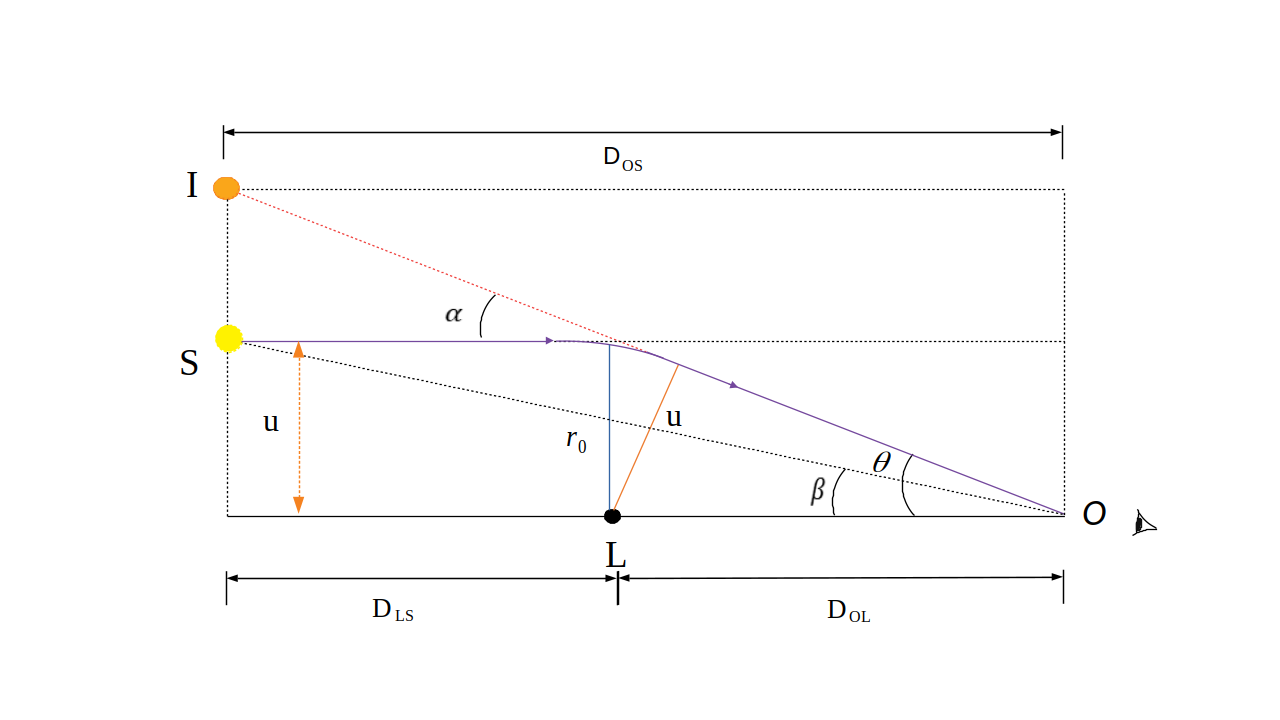}
\caption{ Simplified schematization of the gravitational lensing phenomenon. The light emitted by the source $S$ is slightly deflected from its original trajectory and arrives at the observer $O$ with an angle $\theta$ instead of $\beta$. The deflection angle $\alpha$ is the distance between the image $I$ and the actual position of the source. The relevant distances in the figure are: the impact parameter $u$, the distance of closest approach $x_0$, the distance from the observer to the lens $D_{OL}$ and the distance from the lens to the source $D_{LS}$.}
\label{fig:lentesteste}
\end{figure}

To start with, we show in Fig. \ref{fig:lentesteste} the usual scheme of light deflection, from which we derive the lens equation as \cite{Virbhadra:2002ju,Claudel:2000yi}
\begin{equation}
    \tan \beta = \tan \vartheta- D \left( \tan \vartheta + \tan(\alpha - \theta)  \right ),
    \label{eq:lensequation}
\end{equation}
where \textbf{$\beta$} and $\theta$ are the angular position of the source and the lensed images, respectively, and
\begin{equation}
    D = \frac{D_{LS}}{D_{OS}}
\end{equation}
with $D_{OS}= D_{LS} +D_{OL}$ being the distance between the observer and the source (where $D_{ LS }$ and $D_{OL}$ are the distances marked in the figure).

In order to calculate the angle $\alpha$, we first assume a static and spherically symmetric spacetime defined by
\begin{equation}
    ds^2 = -A(r) dt^2 + B(r) dr^2 + C(r) d \Omega^2,
    \label{eq:metrica1}
\end{equation}
where $d \Omega^2$ is the standard unit sphere metric. 
The deflection angle $\alpha$ is given by \cite{Virbhadra:1998dy}
\begin{equation} 
    \alpha (r_0) = 2 \int_{r_0}^{\infty} \frac{1}{C}  \sqrt{\frac{AB}{1/u^2 -A/C}} dr - \pi. 
    \label{eq:alphageral}
\end{equation}
The above integral can only be solved analytically for some simple cases. Thus, Keeton and Petters \cite{Keeton:2005jd} suggested that this result can be approximated by a series of the following form
\begin{equation}
    \alpha (u) = \mathcal{A}_1 \left(  \frac{m}{u}\right) + \mathcal{A}_2 \left(  \frac{m}{u}\right)^2 + \mathcal{A}_3 \left(  \frac{m}{u}\right)^3 + \mathcal{O} \left(  \frac{m}{u}\right)^4.
    \label{eq:alphau}
\end{equation}
Here, the deflection angle is written as a function of the impact parameter $u$, since it is a gauge invariant variable (while the closest approach distance has a gauge dependence). The $\mathcal{A}_i$ are coefficients to be calculated, which can be simple numbers or depend on a parameter of the solution, such as the charge.
It is worth mentioning that this formalism possesses several limitations, which we consider below.

To demonstrate how this formalism works, we will apply it to the Schwarzschild metric. In this case we have
\begin{equation}
    \begin{aligned}
    & A(r) = 1 - \frac{2 m}{r}, \\
    & B(r) = A(r)^{-1}, \\
    & C(r) = r^2. 
    \end{aligned}
    \label{eq:metricasc}
\end{equation}
Substituting these components into \eqref{eq:alphageral} 
\begin{equation}
    \alpha (r_0) =  2  \int_{r_0}^{\infty}  \frac{1}{r^2}  \frac{dr}{\sqrt{1/u^2 -1/r^2+2m/r^3}}  - \pi,
\end{equation}
to solve this, we first make a coordinate change $x = r_0/r$ and $h = m /r_0$, which leads to the following results
\begin{equation}
    \alpha (r_0) =  2  \int_{0}^{1}    \frac{dx}{\sqrt{1- 2 h -x^2 +2hx^3}}  - \pi.
\end{equation}
Here we used that the relation between $r_0$ and $u$ is
\begin{equation}
    u = \sqrt{\frac{C(r_0)}{A(r_0)}},
    \label{eq:b}
\end{equation}
which in this case becomes
\begin{equation}
    u = \frac{r_0}{\sqrt{1 - \frac{2m}{r_0}}}.
    \label{eq:b1}
\end{equation}
Now, assuming the weak field regime, i.e. $h \ll 1$, we expand the integrand into a Taylor series and then solve the integral term by term, which provides
\begin{eqnarray}
    \alpha(r_0) & = & 4h +\left(\frac{15 \pi }{4}-4\right) h^2 +\left(\frac{122}{3}-\frac{15 \pi }{2}\right) h^3 +\left(\frac{3465 \pi }{64}-130\right) h^4+\left(\frac{7783}{10}-\frac{3465 \pi }{16}\right) h^5 
    \nonumber \\
   &&+ \left(\frac{310695 \pi }{256}-\frac{21397}{6}\right) h^6 + \mathcal{O} (h)^7 \,.
     \label{eq:alphaSCr0}
\end{eqnarray}
To convert this expression into the form \eqref{eq:alphau} we use  
Eq. \eqref{eq:b1} to relate  $r_0$ and $u$ as follows
\begin{equation}
    \frac{r_0}{u} = \frac{2}{\sqrt{3}} \cos \left( \frac{1}{3}  \cos^{-1} \left( - \frac{3^{3/2}m}{u}\right)   \right),
\end{equation}
then, we can write 
\begin{equation}
    r_0 = \frac{4 m}{u}+\frac{15 \pi  m^2}{4 u^2}+\frac{128 m^3}{3 u^3}+\frac{3465 \pi  m^4}{64 u^4}+\frac{3584 m^5}{5 u^5}+\frac{255255 \pi  m^6}{256 u^6}+\mathcal{O}\left( m^7 \right).
    \label{eq:r0euSC}
\end{equation}
Inserting Eq. \eqref{eq:r0euSC} into Eq. \eqref{eq:alphaSCr0}, we finally obtain 
\begin{eqnarray}
    \alpha(u) = 4  \frac{m}{u} + \frac{15 \pi}{4} \left(  \frac{m}{u} \right)^2 + \frac{128}{3} \left(  \frac{m}{u} \right)^3 + \frac{3465 \pi}{4} \left(  \frac{m}{u} \right)^4 +
     \frac{3584}{5} \left(  \frac{m}{u} \right)^5 + \frac{255255 \pi}{256} \left(  \frac{m}{u} \right)^6 + \mathcal{O} \left( \frac{m}{u} \right)^7.
      \label{eq:alphascu}
\end{eqnarray}
Note that with this procedure we obtain Einstein's result, i.e. $\mathcal{A}_1 = 4$, which is twice Newton's result \cite{Weinberg}. Here the coefficients $\mathcal{A}_n$ are only numbers, as the solution has only one parameter, namely the mass $m$.

\subsection{Deflection angle \label{sec:subA1}}

\begin{figure}[htpb]
    \centering
    \includegraphics[width=10cm, height=8cm]{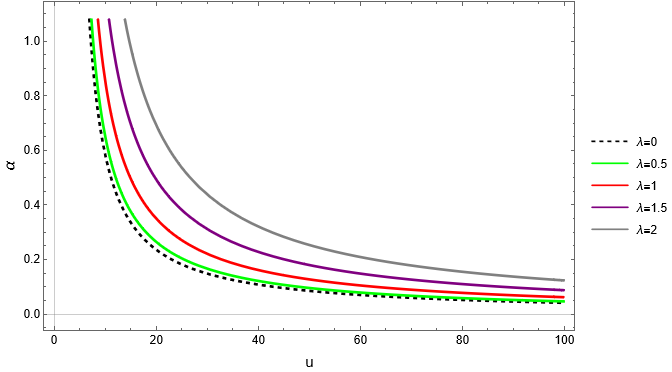}
    \caption{Graphical representation of the deflection angle, given by the equation \eqref{alphauwf} with the coefficients \eqref{eq:coefwf}, for the Schwarzschild with holomy corrections. Here $m=1$, and  $\lambda=0$ represent the standard Schwarzschild solution. }
    \label{fig:alphawf}
\end{figure}

Now, taking into account the the holonomy correction (\ref{Schwholonomy}), which is given by a change in the length scale
\begin{equation}
    l := 2m \frac{\lambda^2}{1+\lambda^2},
\end{equation}
the LQG correction to the Schwarzschild solution (\ref{Schwholonomy}) is given by
\begin{equation}
    \begin{aligned}
    & A(r) = 1 - \frac{2 m}{r}, \\
    & B(r) = \frac{1}{\left(1-\frac{2 m}{r}\right) \left(1-\frac{2m\lambda ^2}{\left(\lambda ^2+1\right) r}\right)}, \\
    & C(r) = r^2. 
    \end{aligned}
    \label{eq:metricah}
\end{equation}
Thus, Eq. \eqref{eq:alphageral} takes the following form 
\begin{equation}
\alpha(r_0) =    2 \int_{r_0}^{\infty} \frac{dr}{r^2} \Biggl[  \frac{\left(\lambda ^2+1\right) r}{\left(\lambda ^2 (r-1)+r\right) \left(\frac{2
   m-r}{r^3}-\frac{2 m}{r_0^3}+\frac{1}{r_0^2}\right)}\Biggr]^{1/2}.
\end{equation}
Note that Eq. \eqref{eq:b1} holds for this model. Now we use the same substitutions as in the previous section, i.e. $x=r_0/r$ and $h = m/r_0$, which leads us to the following results
\begin{equation}
    \alpha(r_0) = 2 \sqrt{\frac{1}{\left(2 h \left(x^3-1\right)-x^2+1\right) \left(1-\frac{2 h \lambda ^2
   x}{\lambda ^2+1}\right)}} - \pi.
\end{equation}
In the weak field regime we have $h\ll 1$ and we can therefore use a Taylor series expansion for the integrand of the above expression in terms of $h$. So we can calculate the integral term by term, which results in
\begin{eqnarray}
    \alpha(r_0) & = & h^5 \left(-\frac{5}{4} (63 \pi -223) \lambda ^2-\frac{3465 \pi }{16}+\frac{7783}{10}\right)+h^4
   \left(\frac{315 \pi  \lambda ^2}{16}-49 \lambda ^2+\frac{3465 \pi }{64}-130\right) 
   \nonumber \\
   && +h^3
   \left(-3 (\pi -5) \lambda ^2-\frac{15 \pi }{2}  +\frac{122}{3}\right)+h^2 \left(\frac{3 \pi 
   \lambda ^2}{2}+\frac{1}{4} \left(15 \pi -8 \left(\lambda ^2+2\right)\right)\right)+2 h
   \left(\lambda ^2+2\right).
\end{eqnarray}
This gives us an approximation for the deflection angle in powers of $m/r_0$. To convert this result into an approximation in terms of $m/u$, we use Eq. \eqref{eq:r0euSC} and get
\begin{equation}
  \begin{aligned}
      \alpha(u) & =  \mathcal{A}_1 \frac{ m}{u} +\mathcal{A}_2 \left( \frac{m}{u} \right)^2 + \mathcal{A}_3    \left( \frac{m}{u} \right)^3 
  +\mathcal{A}_4  \left( \frac{m}{u} \right)^4  + \mathcal{A}_5  \left( \frac{m}{u} \right)^5 +  \mathcal{O}\left( \frac{m}{u} \right)^6 ,
  \end{aligned}
  \label{alphauwf}
\end{equation}
where
\begin{equation}
    \begin{aligned}
         \mathcal{A}_1 = 2 \left(\lambda ^2+2\right), \ \ \ \mathcal{A}_2 =\frac{3 \pi  \left(2 \lambda
   ^2+5\right) }{4 },    \ \ \ \mathcal{A}_3 =  \frac{16 \left(3 \lambda ^2+8\right)}{3 },
   \ \ \ \mathcal{A}_4 =\frac{315 \pi  \left(4 \lambda ^2+11\right)
  }{64 }, \ \ \ \mathcal{A}_5 = \frac{256 \left(5 \lambda ^2+14\right)}{5}.
    \end{aligned}
    \label{eq:coefwf}
\end{equation}
This result shows that the constant polymerization parameter $\lambda$ increases the deflection angle. One can  easily  verify that if $\lambda = 0$, then Eq. \eqref{eq:alphascu} is recovered.
However, in this case we have $\mathcal{A}_1 \neq 4$, which emphasizes the fact that we are not dealing with classical general relativity.
In Fig. \ref{fig:alphawf}, we plot $\alpha$ against the impact parameter $u$.

\subsection{Positions of lensed images \label{sec:subA2}}

From now on, we use the coefficients of the deflection angle expansion given in Eq. \eqref{alphauwf} to calculate the following observables: the image position, the magnification and the time delay between the  primary and secondary images. First, we change the variables in the lens equation \eqref{eq:lensequation} by using the weak Einstein deflection ring radius,
\begin{equation}
    \theta_{E} = \sqrt{\frac{4 G m D_{LS}}{c^2D_{OL}D_{OS}}} = \sqrt{\frac{4  m D_{LS}}{D_{OL}D_{OS}}} \,
    \label{eq:thetaE}
\end{equation}
and we define
\begin{equation}
  \beta  = \frac{\mathfrak{\beta}}{\theta_{E}}, \qquad \theta = \frac{\vartheta}{\theta_{E}}, \qquad \epsilon =\frac{\tan^{-1}(m/D_{OL})}{\theta_{E}} = \frac{\theta_{E}}{4 D}\,.
    \label{eq:novosangulos}
\end{equation}
We now assume a solution of the lens equation \eqref{eq:lensequation} in the form (the Einstein ring is a natural scale in this context, so we will write all the other quantities involved in $\epsilon$ power expansions)
\begin{equation}
    \theta = \theta_0 + \theta_1 \epsilon + \theta_2 \epsilon^2 + \theta_3 \epsilon^3 + \mathcal{O} (\epsilon)^4 \,.
\end{equation}
Then we can write the deflection angle as 
\begin{equation}
    \alpha = \frac{\mathcal{A}_1}{\theta_0} \epsilon + \frac{\mathcal{A}_2 - \mathcal{A}_1\theta_1}{\theta_0^2}\epsilon^2 + \frac{1}{\theta_0^3} \left[ \mathcal{A}_2 - 2 \mathcal{A}_3\theta_1 + \mathcal{A}_1 \left(  \frac{8}{3}D^2\theta_0^4 +\theta_1^2 - \theta_0\theta_2  \right)  \right]\epsilon^3 + \mathcal{O} (\epsilon)^4.
\end{equation}

\begin{figure}[htpb]
    \centering
   \includegraphics[width=12cm, height=10cm]{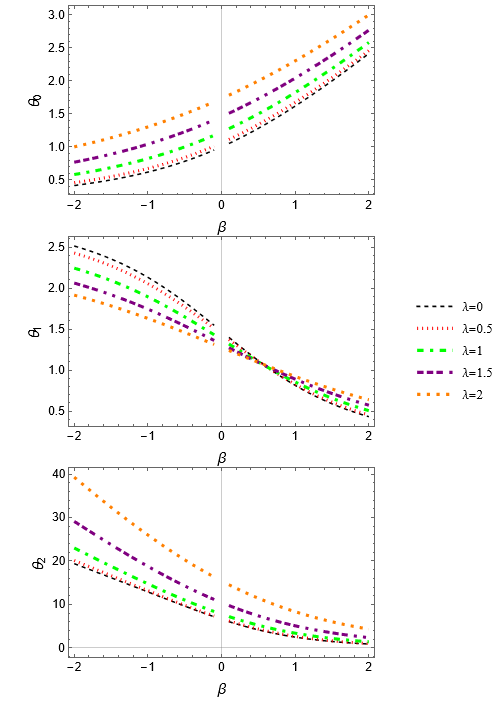}
    \caption{Graphical representation of the components of the $\theta$ expansion, biven by the equation \eqref{eq:thetacoef}, as a function of the angle $\beta$. We have considered $D = 0.01$.
 We have the positive-parity image when $\beta >0$
and negative-parity image for $\beta <0$. We omit the regions with $|\beta| =0.1$ for obvious physical reasons. }
    \label{fig:theta012}
\end{figure}

Putting these pieces together, we fix $\beta$ and solve the lens equation term by term to find the coefficients $\theta_i$ \cite{Keeton:2005jd}. The first one is
\begin{equation}
   \theta_0 =  \frac{1}{2} \left(\sqrt{\beta ^2+\mathcal{A}_1}+\beta \right)\,.
   \label{eq:theta0}
\end{equation}
The other coefficients are given in terms of $\theta_0$ and $\mathcal{A}_i$. For the solution considered we have
\begin{equation}
    \theta_1 = \frac{3 \pi  \left(2 \lambda ^2+5\right)}{8 \left(2 \theta_0^2+\lambda ^2+2\right)},
\end{equation}
\begin{equation}
    \begin{aligned}
        \theta_2 = & \Biggl\{192
   \theta_0 \left(2 \theta_0^2+\lambda ^2+2\right)^3\Biggr\}^{-1}    \Biggl\{512 D^2 \Bigl[4 \theta_0^8 \left(\lambda ^2-10\right)+4 \theta_0^6 \left(7 \lambda ^4+16 \lambda ^2+16\right)+\theta_0^4 \left(7 \lambda ^6+30
   \lambda ^4+84 \lambda ^2+88\right) 
   \\ &
   +4 \theta_0^2 \left(3 \lambda ^4+8 \lambda
   ^2+4\right)-4 \left(\lambda ^2+2\right)^2\Bigr] 
   -1536 D \theta_0^2 \left(\lambda ^2+2\right)^2 \left(2 \theta_0^2+\lambda ^2+2\right)^2+256 \bigl[4
   \theta_0^4 \left(\lambda ^6+6 \lambda ^4+18 \lambda ^2+24\right) 
   \\ &
   + \theta_0^2 \left(\lambda ^2+2\right) \left(4 \lambda ^6+24 \lambda ^4+69 \lambda
   ^2+90\right)+\left(\lambda ^2+2\right)^2 \left(\lambda ^6+6 \lambda ^4+18 \lambda
   ^2+24\right)\bigr]-27 \pi ^2 \left(\lambda ^2+2\right) \left(2 \lambda ^2+5\right)^2 \Biggr\}.
    \end{aligned}
    \label{eq:thetacoef}
\end{equation}
The first order correction of the position of the images in the order of $\epsilon$ is thus
\begin{equation}
    \theta = \theta_0 + \frac{3 \pi  \left(2 \lambda ^2+5\right)}{8 \left(2 \theta_0^2+\lambda ^2+2\right)} \epsilon + \mathcal{O}( \epsilon)^2.
\end{equation}
We can see that there is no influence of the $\lambda$ on the zero order, only from the first upwards. In Fig. \ref{fig:theta012}, we represent the coefficients $\theta_{0,1,2}$ as functions of the angle $\beta$ (the angles defined by Eq. \eqref{eq:novosangulos}) for different values of $\lambda$ . We choose $D= 0.01$  inspired by real situations in which $D_{OS} \gg D_{LS}$.

\subsection{Magnifications \label{sec:subA3}}

The optical magnification $\mu$, how much an image has decreased or increased in apparent size, is defined by
\begin{equation}
    \mu = \Biggl[ \frac{\sin(\beta)}{\sin (\theta)} \frac{d \beta}{d\theta} \Biggr]^{-1},
\end{equation}
at a angular position $\theta$.  As before, we can construct a series expansion for the magnification in terms of $\epsilon$
\begin{equation}
    \mu = \mu_0 + \mu_1 \epsilon +\mu_2 \epsilon^2 + \mu_3 \epsilon^3 + \mathcal{O}( \epsilon)^4.
\end{equation}

For the Schwarzschild with holonomy corrections solution we have
\begin{eqnarray}
    \mu_0 = \frac{16 \theta_0^4}{16 \theta_0^4-4 \left(\lambda ^2+2\right)^2},
    \qquad 
     \mu_1 = -\frac{3 \pi  \theta_0^3 \left(2 \lambda ^2+5\right)}{2 \left(2 \theta_0^2+\lambda ^2+2\right)^3},
\end{eqnarray}
\begin{equation}
    \begin{aligned}
        \mu_2 &= \Biggl\{  3 \left(2 \theta_0^2+\lambda ^2+2\right)^5 \left(4
   \theta_0^2-2 \left(\lambda ^2+2\right)\right)^2\Biggr\}^{-1}    \Biggl\{    
   4 \theta_0^2 \Bigl[-8 D^2 \left(\lambda ^2+2\right)^2 \left(2
   \theta_0^2+\lambda ^2+2\right)^2 \bigl(-18 \left(\theta_0^2-3\right)
   \lambda ^4
   \\ &+36 \left(\theta_0^4-2 \theta_0^2+3\right) \lambda ^2-8
   (\theta_0-1) (\theta_0+1) \left(\theta_0^4+16 \theta_0^2+1\right)+9 \lambda ^6\bigr) -192 D \theta_0^2 \left(\lambda
   ^2+2\right)^3 \left(2 \theta_0^2-\lambda ^2-2\right) \left(2 \theta_0^2+\lambda ^2+2\right)^2 \\ &-\frac{1}{2} \theta_0^2 \left(2 \theta_0^2-\lambda ^2-2\right) \bigl(512 \theta_0^4 \left(\lambda ^6+6 \lambda ^4+18
   \lambda ^2+24\right)+\theta_0^2 \left(512 \left(\lambda ^2+2\right) \left(\lambda
   ^6+6 \lambda ^4+18 \lambda ^2+24\right)-81 \pi ^2 \left(2 \lambda ^2+5\right)^2\right)
   \\&+128 \left(\lambda ^2+2\right)^2 \left(\lambda ^6+6 \lambda ^4+18 \lambda
   ^2+24\right)\bigr)\Bigr]\Biggr\}.
    \end{aligned}
\end{equation}
We use the expressions for the general case in \cite{Keeton:2005jd}, since $\mathcal{A}_1 \neq 4$. Note that in this case the parameter $\lambda$ influences already from order zero.

\subsection{Total Magnification and Centroid \label{sec:subA4}}

Taking into account \eqref{eq:theta0} and considering that $\beta$ can also take negative values, we have
\begin{equation}
       \theta_0^{\pm} =  \frac{1}{2} \left(\sqrt{\beta ^2+\mathcal{A}_1}\pm |\beta| \right) \,.
\end{equation}
This means that $\theta_0^{+}$ is an image that is on the same side as the source and the lens, and $\theta_0^{-}$ is an image that is on the opposite side of both, so that the magnification also becomes $\mu^{\pm}$. In observations known as microlensing, two or more images cannot be resolved together. What we observe then is the total magnification and the magnification-weighted centroid. As described in \cite{Keeton:2005jd}, these observables have no first order corrections for classical general relativity because $\mathcal{A}_1= 4 $ makes this term zero. For our case this does not occur and we have a non-zero first order term. Therefore, the total magnification $\mu_{t} = |\mu^{+}| + |\mu^{-}|$, can be rearranged to the following expression
\begin{eqnarray}
    \mu_{t} &=&
   -\frac{3 \pi  \theta_0^3 \left(\theta_0^2-1\right)
   \lambda ^2 \left(2 \lambda ^2+5\right)   \left(6 \left(\theta_0^2+1\right)^2 \lambda ^2+12 \left(\theta_0^2+1\right)^2+\left(\theta_0^4+\theta_0^2+1\right) \lambda ^4\right)}{2 \left(2 \theta_0^2+\lambda
   ^2+2\right)^3 \left(\theta_0^2 \left(\lambda ^2+2\right)+2\right)^3}\epsilon
   \nonumber \\
&& +  \frac{4 \left(\theta_0^8-1\right) \left(\lambda ^2+2\right)^2}{\left(4 \theta_0^4-\left(\lambda ^2+2\right)^2\right) \left(\theta_0^4 \left(\lambda
   ^2+2\right)^2-4\right)} \,.
   \label{eq:mucoef}
\end{eqnarray}

\begin{figure}[htpb]
    \centering
   \includegraphics[width=12cm, height=10cm]{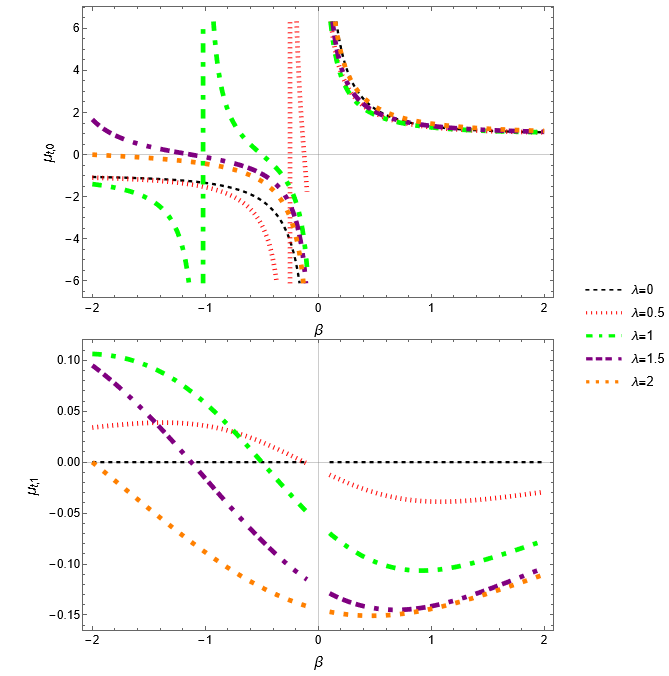}
    \caption{Graphical representation of the total magnification, given by the equation \eqref{eq:mucoef}, as a function of the angle $\beta$. We considered $\epsilon = 1$.  We have the positive-parity image when $\beta >0$
and negative-parity image for $\beta <0$. We omit the regions with $|\beta| =0.1$ for obvious physical reasons.}
    \label{fig:mutotal}
\end{figure}

We can use the Eq. \eqref{eq:theta0} to give the total magnification as a function of the angle $\beta$, but the resulting expression is extremely lengthy, so we do not reproduce it here.  In Fig. \ref{fig:mutotal}, we show the total magnification highlighting each order $\mu_{t,0}$ and $\mu_{t,1}$ as a function of the angle $\beta$. Note that $\mu_{t,1}= 0$ for $\lambda=0$, which would correspond to the result of classical general relativity. We made $\epsilon=1$  because we just want to analyze the graphical behavior and  of the loop quantum gravity parameter, in observational situations $\epsilon \sim 10^{-4}$.

\begin{figure}[htpb]
    \centering
   \includegraphics[width=12cm, height=10cm]{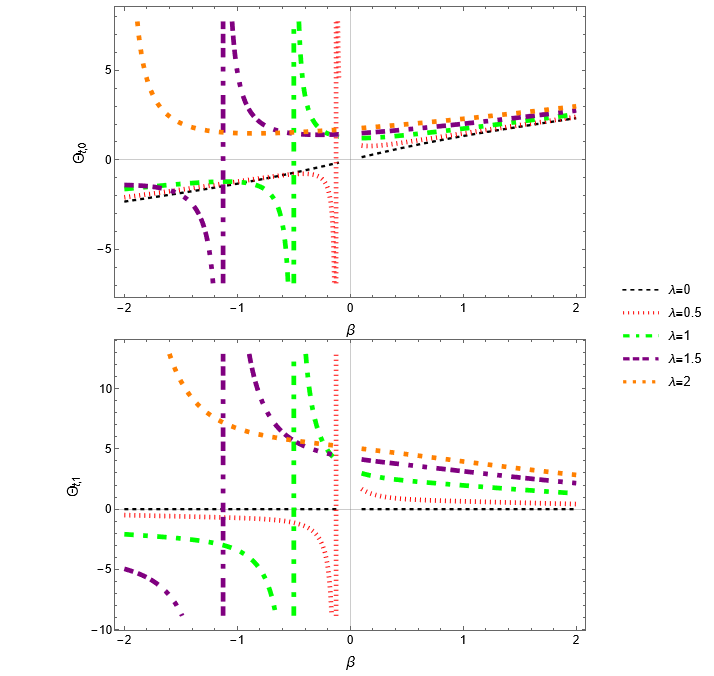}
    \caption{Graphical representation of the magnification-weighted centroid, given by the equation \eqref{eq:centcoef}, as a function of the angle $\beta$, where we consider $\epsilon = 1$.  We have the positive-parity image when $\beta >0$
and negative-parity image for $\beta <0$. We omit the regions with $|\beta| =0.01$ for obvious physical reasons.}
    \label{fig:centroid}
\end{figure}

The magnification-weighted centroid is defined by
\begin{equation}
    \Theta = \frac{\theta^{+}|\mu^{+}|-\theta^{-}|\mu^{-}|}{|\mu^{+}|+ |\mu^{-}|},
\end{equation}
which leads to 
\begin{equation}
    \begin{aligned}
       \Theta =& \frac{4 \left(\theta_0^8-\theta_0^6+\theta_0^4-\theta_0^2+1\right) \left(\lambda ^2+2\right)^2-16 \theta_0^4}{4 \theta_0
   \left(\theta_0^6-\theta_0^4+\theta_0^2-1\right) \left(\lambda
   ^2+2\right)^2} 
    +\epsilon  \Biggl\{3 \pi  \theta_0^2 \lambda ^2 \left(2 \lambda ^2+5\right)  \times
   \\
  & \times \Biggl[\frac{1}{4} \left(\theta_0^{12}+1\right) \left(\lambda ^2+2\right)^2 \left(9
   \pi ^2 \left(2 \lambda ^2+5\right)^2+128 \left(\lambda ^2+4\right)\right)+16 \theta_0^6 \left(\lambda ^8+20 \lambda ^6+104 \lambda ^4+208 \lambda ^2+160\right) \\&
   +16 \left(\theta_0^4+1\right) \theta_0^4 \left(\lambda ^8+12 \lambda ^6+40
   \lambda ^4+40 \lambda ^2+16\right)+32 \left(\theta_0^8+1\right) \theta_0^2 \left(\lambda ^2+2\right) \left(\lambda ^4+4 \lambda ^2-4\right)\Biggr]\Biggr\} \times \\
  & \left\{128 \left(\theta_0^4-1\right) \left(\theta_0^4+1\right)^2
   \left(\lambda ^2+2\right)^4 \left(2 \theta_0^2+\lambda ^2+2\right)
   \left(\theta_0^2 \left(\lambda ^2+2\right)+2\right)\right\}^{-1} \,.
    \end{aligned}
    \label{eq:centcoef}
\end{equation}
The expression of this observable as a function of the angle $\beta$ is also too extensive and we have therefore decided not to write it here. However, we show in the Fig. \ref{fig:centroid}
the magnification-weighted centroid as a function of this angle for different values of $\lambda$. We note that, again, the first-order component is zero when $\lambda=0$.

\subsection{Time Delay}\label{sec:subA5}

\begin{figure}[htpb]
    \centering
   \includegraphics[width=12cm, height=10cm]{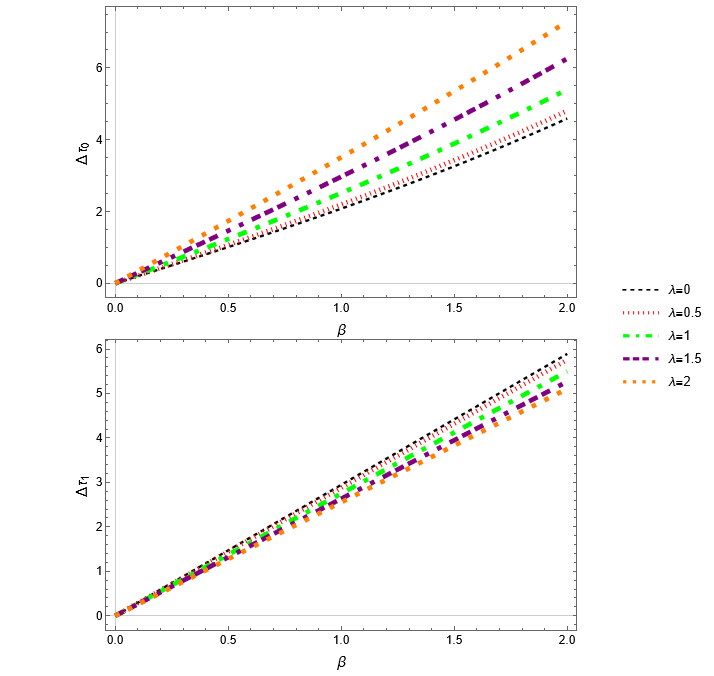}
    \caption{Graphical representation of the time delay coefficients, given by the equation \eqref{eq:taucoef}, as a function of the angle $\beta$. }
    \label{fig:delaywf}
\end{figure}

The path followed by the photons of the first image with positive parity and the second image with negative parity is different and with this we can calculate the associated delay time.
We can write the delay time and the first-order correction in the form \cite{Lukmanova:2018dwz}

\begin{equation}
    \Delta \tau = \Delta \tau_0 + \Delta \tau_1 \epsilon + \mathcal{O} (\epsilon)^2,
\end{equation}
where
\begin{equation}
    \begin{aligned}
        & \Delta \tau_0 = \frac{1}{2}|\beta| \sqrt{\mathcal{A}_1+\beta^2} +\frac{\mathcal{A}_1}{4} \ln \left(\frac{\sqrt{\mathcal{A}_1+\beta^2}+\beta }{\sqrt{\mathcal{A}_1+\beta^2}-\beta} \right), \\
   & \Delta \tau_1 = \frac{\mathcal{A}_2}{\mathcal{A}_1} |\beta|.
    \end{aligned}
    \label{eq:delayg}
\end{equation}
From Eq. \eqref{eq:delayg} it is clear that at $\beta =0$ (source, lens and observer are aligned) we have the Einstein ring and there is no delay between images.
If we now consider the coefficients given by Eq. \eqref{eq:coefwf}, we get that
\begin{equation}
    \begin{aligned}
        & \Delta \tau_0 = \frac{1}{2} \beta  \sqrt{\beta ^2+2 \left(\lambda ^2+2\right)}+\frac{1}{2} \left(\lambda ^2+2\right) \ln
   \left(\frac{\sqrt{\beta ^2+2 \left(\lambda ^2+2\right)}+\beta }{\sqrt{\beta ^2+2 \left(\lambda
   ^2+2\right)}-\beta }\right), \\
   & \Delta \tau_1 = \frac{3 \pi  \beta  \left(2 \lambda ^2+5\right)}{8 \left(\lambda ^2+2\right)},
    \end{aligned}
    \label{eq:taucoef}
\end{equation}
this means that the lambda parameter increases the delay time. In Fig. \ref{fig:delaywf} we show the influence of $\beta$ on the coefficients $\Delta \tau_{0,1}$ for different values of $\lambda$.

\section{Deflection angle by Gauss-Bonnet theorem} \label{sec:GB}

In this section we will re-derive the coefficient $\mathcal{A}_1$ for the deflection angle in the weak field regime using the method described in \cite{Gibbons:2008rj}.
Its equivalence with the geodesic method was shown in \cite{Li:2019mqw}. In summary, Gibbons and Werner proposed that the Gauss-Bonnet theorem 
can be used to calculate the deflection angle in the form
\begin{equation}
    \alpha = - \int_{0}^{\pi}\int_{r_{sl}}^{\infty}  K dS.
    \label{eq:GWformula}
\end{equation}
Here $K$ is  the optical Gaussian
curvature and $r_{sl}$ is the distance traveled by the photon considering the approximation of a straight line path. 
We consider, without loss of generality, that the movement takes place in the equatorial plane where $\theta = \pi/2$, so we have
\begin{equation}
    ds^2 = -A(r) dt^2 + B(r)dr^2 + C(r) d \phi^2 \,.
\end{equation}
From that metric we derive the optical metric $\Bar{g}_{ij}$ from the relation
\begin{equation}
    dt^2 = \Bar{g}_{rr} dr^2 + \Bar{g}_{\phi \phi} d \phi^2 = \frac{B(r)}{A(r)} dr^2 + \frac{C(r)}{A(r)} d \phi^2.
\end{equation}
We assume that the geodesics of photons obey $ds^2= 0$. The Gaussian curvature $K$ is defined in terms of the new metric $\Bar{g}_{ij}$ as follows
\begin{equation}
    K = -\frac{1}{\sqrt{\Bar{g}_{rr}\Bar{g}_{\phi \phi}}} \Biggl[  \frac{\partial }{\partial r}  \left( \frac{1}{\sqrt{\Bar{g}_{rr}}} \frac{\partial \sqrt{\Bar{g}_{\phi \phi}}}{\partial r} \right) +
    \frac{\partial}{\partial \phi} \left(  \frac{1}{\sqrt{\Bar{g}_{\phi \phi}}} \frac{\partial \sqrt{\Bar{g}_{rr}}}{\partial \phi}    \right) \Biggr].
    \label{eq:Kgeral}
\end{equation}
In addition, we have that
\begin{equation}
    dS^2 = \sqrt{\text{det} |\Bar{g}|} dr d\phi,
\end{equation}
and
\begin{equation}
    r_{sl} = \frac{u}{\sin \phi}.
\end{equation}

Let us now apply this formalism and calculate the first coefficient of the deflection angle expansion in the weak field regime. From the Eq. \eqref{eq:metricah} we have
\begin{equation}
    \Bar{g}_{r r} =\frac{1}{\left(1-\frac{2 m}{r}\right)^2 \left(1-\frac{2 \text{m$\lambda $}^2}{\left(\lambda ^2+1\right)
   r}\right)} , \ \ \ \Bar{g}_{\phi \phi} =\frac{r^2}{1-\frac{2 m}{r}} .
   \label{eq:gbar}
\end{equation}
Substituting Eq. \eqref{eq:gbar} into Eq. \eqref{eq:Kgeral} and considering a Taylor series expansion up to the order of $\lambda^2$ we get  
\begin{equation}
    \alpha = - \int_{0}^{\pi}\int_{\frac{u}{\sin \phi}}^{\infty} \Biggl[ \frac{m (3 m-2 r)}{(r-2 m)^3 \left(\frac{r}{r-2 M}\right)^{3/2}}
 +  \frac{\lambda ^2 m \left(9 m^2-7 m r+r^2\right) \left(-\frac{r}{2 M-r}\right)^{3/2}}{r^4} + \mathcal{O} (\lambda^4) \Biggr] dr d \phi.
 \label{eq:gbalpha}
\end{equation}
The first integral, in the coordinate $r$, results in
\begin{equation}
     \alpha = - \int_{0}^{\pi} \Biggl[1 -\frac{\lambda ^2 m \sin (\phi ) (u-3 m \sin (\phi )) \sqrt{\frac{u}{u-2 m \sin (\phi )}}}{u^2}+\frac{3 m
   \sqrt{\frac{1}{1-2 \sin (\phi )}} \sin (\phi )}{u}-\sqrt{\frac{u}{u-2 m \sin (\phi )}}
     \Biggr]d \phi.
\end{equation}
As we are dealing with the weak field regime, where $m/u \ll 1$, we rewrite the above integral considering a Taylor series expansion as follows
\begin{equation}
     \alpha = - \int_{0}^{\pi} \Biggl[ -\frac{3 m \sin(\phi ) \left(\frac{m \sin (\phi )}{u}+1\right)}{u}-\frac{\lambda ^2 m \sin (\phi
   )}{u}+\frac{m \sin (\phi )}{u} + \mathcal{O} \left( \frac{m}{u} \right)^2  \Biggr]d \phi.
   \label{eq:50}
\end{equation}
The result is then
\begin{equation}
    \alpha =2 (2 + \lambda^2)\frac{m}{u}.
\end{equation}
Therefore, we recover the same $\mathcal{A}_1$ found by the geodesics method \eqref{eq:coefwf} in section \ref{sec:one}.

\section{Strong gravitational lensing} \label{sec:two}

 In the previous section, we used a formalism that applies to the case where the closest approach ($r_0$) distance is much larger compared to the mass of the lens (in our case, the black hole). In this section, we will discuss the deflection of light in the so-called strong field regime.  From the point of view of the classical mechanics we would expect that the light (or the particles) can approach the event horizon freely and would be in an "inaccessible" region only after crossing it. But this is not what actually happens. In fact, the light is absorbed by the black hole when it is sent with a radius smaller than the critical impact parameter ($u_m$) of the solution. K. Virbhadra and G. Ellis \cite{Virbhadra:1999nm} studied the gravitational lensing effect for the Schwarzschild solution and showed that it occurs for the value $r_0 < 3 \sqrt{3} m$. They also showed that $r= 3 m$ delimits a surface on which any null geodesic starting at any point on the surface and initially tangent to it remains in the same surface, which we call the photon sphere. In any spacetime containing a photonsphere, gravitational lensing leads to relativistic images \cite{Virbhadra:1999nm}. As $r_0 \rightarrow r_m$ (or $u \rightarrow u_m$), the deflection angle increases and consequently diverges to $r_0=r_m$ (or $u=u_m$).
This concept of the photon surface of the Schwarzschild solution was generalized in \cite{Claudel:2000yi}, we can calculate $r_m$ by
\begin{equation}
    \frac{C^{\prime}(r)}{C(r)} = \frac{A^{\prime}(r)}{A(r)}.
    \label{eq:fotonesfera}
\end{equation}
So, in this section we study the deflection of light in the limit where the geodesics pass close to the photon surface. 

\subsection{Deflection Angle}

Considering that the deflection angle tends to infinity as $r_0 \rightarrow r_m $,  Bozza \cite{Bozza:2002zj} proposed that in this limit $\alpha$ can be approximated by a logarithmic expansion of the form
\begin{equation}
    \alpha (u) = b_1 \log \left( \frac{u}{u_m} - 1 \right) + b_2 + \mathcal{O} \left( u - u_m \right) \,,
    \label{eq:explog}
\end{equation}
where $b_1$, $b_2$ and $u_m$ are coefficients to be calculated from the metric. The critical impact parameter $u_m$ is obtained directly from Eq. \eqref{eq:b} if we equate the distance of closest approach $x_0$ with the radius of the photon sphere $x_m$.
We will briefly discuss how to obtain the other two coefficients. We emphasize that the starting point is the same as before, namely to propose an approximate result for the integral \eqref{eq:alphageral}. First, we assume the metric (we use the substitution $x= r/2m$)
\begin{equation}
   ds^2 = -A(x) dt^2 + B(x) dr^2 + C(x) d \Omega^2,
\end{equation}
Then, we define the variables
\begin{equation}
    y = A(x),
\end{equation}
\begin{equation}
    \zeta = \frac{y - y_0}{1 - y_0},
\end{equation}
where $y_0 = A(x_0)$.  This leads to
\begin{equation}
    \alpha (x_0) = I(x_0) - \pi,
\end{equation}
\begin{equation}
    I(x_0) = \int_0^1 R( \zeta, x_0) f(\zeta, x_0) d \zeta.
    \label{eq:Itotal}
\end{equation}
Here the function $R( \zeta, x_0)$ is given by
\begin{equation}
    R( \zeta, x_0) = \frac{2 \sqrt{By}}{C A^{\prime}} \left( 1 - y_0 \right) C_0, 
\end{equation}
which is regular for any value of $\zeta$ and $x_0$. The function $f(\zeta, x_0)$ is provided by
\begin{equation}
    f(\zeta, x_0) = \frac{1}{\sqrt{y_0 - \bigl[ (1-y_0) \zeta + y_0\bigr]\frac{C_0}{C}}},
    \label{eq:f01}
\end{equation}
which has a divergence for $\zeta \rightarrow 0$.  All functions without the subscript 0 are evaluated
at $x = A^{-1}\bigl[(1 - y_0) \zeta + y_0 \bigr]$. We rewrite $f(\zeta, x_0)$  as
\begin{equation}
    f(\zeta, x_0) \sim f_0(\zeta, x_0) = \frac{1}{\sqrt{\beta_1 \zeta + \beta_2 \zeta^2}},
    \label{eq:f02}
\end{equation}
where
\begin{equation}
    \beta_1 = \frac{1 - y_0}{C_0 A^{\prime}_0} \left(  C^{\prime}_0 y_0 - C_0 A^{\prime}_0  \right),
    \label{eq:beta1}
\end{equation}
and
\begin{equation}
    \beta_2 = \frac{(1-y_0)^2}{2C_0^2 A^{\prime \ 3}_0}  \bigr[  2 C_0 C^{\prime}_0 A^{\prime \ 2}_0 +(C_0 C_0^{\prime \prime} - 2C_0^{\prime \ 2} )y_0 A_0^{\prime} - C_0 C_0^{\prime}y_0A_0^{\prime \prime}  \bigl].
    \label{eq:beta2}
\end{equation}
In the form \eqref{eq:f02} we can see that: If $\beta_1 \neq 0$, the leading order of divergence in \eqref{eq:f01} is $\zeta^{-1/2}$, and if $\beta_1 = 0$ the divergence is $\zeta^{-1}$. In the first case $f_0$ can be integrated and the result is finite, while in the second case the integral diverges. Returning to the original variables, we note that $\beta_1$ vanishes at $x_0 = x_m$, and in order to solve this problem, we treat the integral \eqref{eq:Itotal} as follows
\begin{equation}
    I(x_0) = I_D (x_0) + I_R (x_0),
\end{equation}
where
\begin{equation}
    I_D (x_0) = \int_0^1 R( 0, x_m) f_0(\zeta, x_0) d \zeta,
\end{equation}
refers to the divergent part, and
\begin{equation}
    I_R (x_0) = \int_{0}^1 g(\zeta, x_0) d \zeta,
\end{equation}
with
\begin{equation}
    g(\zeta, x_0) = R( \zeta, x_0) f(\zeta, x_0) - R( 0, x_m) f_0(\zeta, x_0).
\end{equation}
Note that $I_R = I - I_D$. The result of these integrals is \cite{Bozza:2002zj}
\begin{equation}
    I_D(x_0) = - \left( \frac{R(0,x_m)}{\sqrt{\beta_{2m}}}\right) \log \left( \frac{x_0}{x_m} -1 \right) +  \frac{R(0,x_m)}{\sqrt{\beta_{2m}}} \log \frac{2(1- y_m)}{A^{\prime}_m x_m} + \mathcal{O} \left( x_0 - x_m\right),
\end{equation}
and
\begin{equation}
    I_R(x_m) = \int_0^1 g(\zeta, x_m) d \zeta + \mathcal{O} \left( x_0 - x_m\right).
\end{equation}
Functions with index $m$ are calculated in  $x_0 = x_m$.  The logarithmic approximation for the deflection angle is then 
\begin{equation}
    \alpha (x_0) =  - \left( \frac{R(0,x_m)}{\sqrt{\beta_{2m}}}\right) \log \left( \frac{x_0}{x_m} -1 \right) +  \frac{R(0,x_m)}{\sqrt{\beta_{2m}}} \log \frac{2(1- y_m)}{A^{\prime}_m x_m} +\int_0^1 g(\zeta, x_m) d \zeta- \pi + \mathcal{O} \left( x_0 - x_m\right).
   \label{eq:alphax0strong}
\end{equation}
As stated in the previous section, it is convenient to write this result in terms of the gauge invariant coordinate $u$. We can expand Eq. \eqref{eq:b} and write
\begin{equation}
    u - u_m = \beta_{2m} \sqrt{\frac{y_m}{C_m^3}} \frac{C^{\prime \ 2}_m }{2(1-y_m^2)} \left( x_0 - x_m \right)^2.
    \label{uex0}
\end{equation}
With the above equation we can write Eq. \eqref{eq:alphax0strong} in the form \eqref{eq:explog}, where the coefficients $b_1$ and $b_2$ are
\begin{equation}
    b_1 = \frac{R(0,x_m)}{2\sqrt{\beta_{2m}}},
    \label{eq:b1g}
\end{equation}
\begin{equation}
    b_2 = \int_0^1 g(\zeta, x_m) d \zeta +  b_1 \log \frac{2 \beta_{2m}}{y_m} - \pi.
    \label{eq:b2g}
\end{equation}

\begin{figure}[htpb]
    \centering
   \includegraphics[width=10cm, height=8cm]{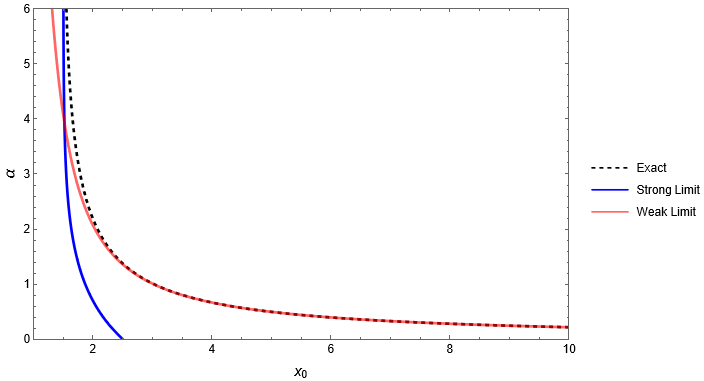}
    \caption{Graphical representation of the deflection angle for the Schwarzschild solution.}
    \label{fig:comp}
\end{figure}

Let's now apply this mechanism to the  Schwarzschild solution. In standard coordinates, the metric is
\begin{equation}
    \begin{aligned}
    & A(x) = 1- \frac{1}{x}, \\
    & B(x) = A(x)^{-1}, \\
    & C(x) = x^2. 
    \end{aligned}
    \label{eq:SCstandart}
\end{equation}
The first step is to find the radius of the photon sphere $x_m$ from Eq. \eqref{eq:fotonesfera}, which in this case is
\begin{equation}
    x_m = \frac{3}{2},
    \label{eq:xmsc}
\end{equation}
 Substituting this result into Eq. \eqref{eq:b}, we immediately get $u_m = 3 \sqrt{3}/2$.
The functions $R(\zeta, x_0)$ and $f(\zeta, x_0)$ are given by
\begin{equation}
    \begin{aligned}
       R(\zeta,x_0) &= 2, \\
       f(\zeta, x_0) &= \frac{1}{\sqrt{\beta_1 \zeta +\beta_2 \zeta^2- \frac{\zeta^3}{x_0}}},
       \label{eq:fersc}
    \end{aligned}
\end{equation}
respectively, where, from Eqs. \eqref{eq:beta1} and \eqref{eq:beta2}, we have
\begin{equation}
    \begin{aligned}
        & \beta_1 = 2 - \frac{3}{x_m} =0, \\
        & \beta_2 =  \frac{3}{x_m} -1 =1.
    \end{aligned}
    \label{eq:beta12}
\end{equation}
and with these results we find $b_1=1$. The integral $I_R$ is
\begin{equation}
    I_R(x_m) = \int_0^1 g(\zeta, x_m) d \zeta =  0.9496,
\end{equation}
then $b_2= -0.4002$. Putting these results together, we get the logarithmic expansion for the deflection angle as a function of the impact parameter, given by
\begin{equation}
    \alpha (u) = \log \left( \frac{2u}{3 \sqrt{3}} - 1 \right) -0.4002.
\end{equation}
In Fig. \ref{fig:comp} we show the exact deflection angle (computed numerically with the integral \eqref{eq:alphageral}) and the strong/weak field approximations.
We see that both the logarithmic expansion (blue line) and the exact values (dotted line) diverge at $x_0 \rightarrow x_m = 1.5$, but these curves move away rapidly as $x_0$ increases. In contrast, the weak-field expansion approaches the exact result as $x_0$ increases.

\begin{figure}[htpb]
    \centering
   \includegraphics[width=10cm, height=8cm]{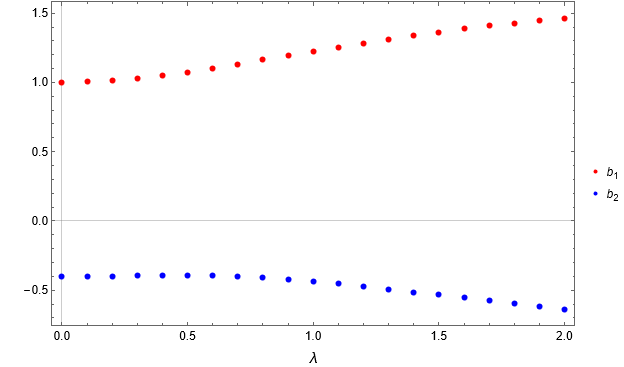}
    \caption{Graphical representation, for the Schwarzschild solution with holonomy corrections, of the coefficients as a function of the  parameter $\lambda$.}
    \label{fig:coef}
\end{figure}

\begin{figure}[htpb]
    \centering
   \includegraphics[width=10cm, height=8cm]{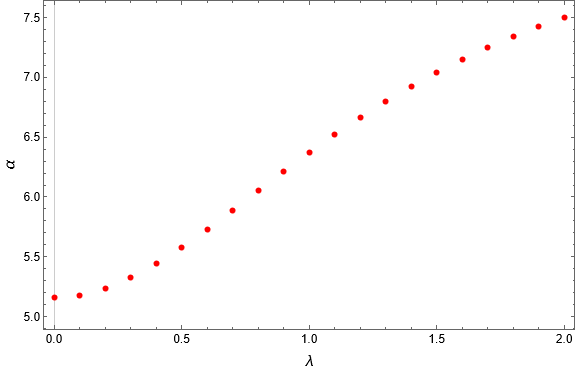}
    \caption{Graphical representation, for the Schwarzschild solution with holonomy corrections, of the deflection angle as a function of the  parameter $\lambda$.}
    \label{fig:alphalambda}
\end{figure}

\begin{figure}[htpb]
    \centering
   \includegraphics[width=10cm, height=8cm]{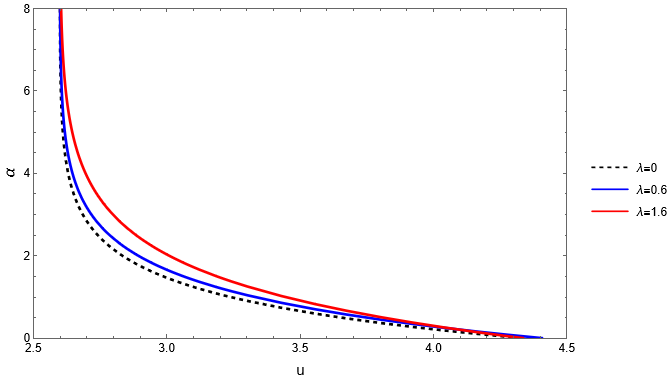}
    \caption{Graphical representation, for the Schwarzschild solution with holonomy corrections, of  the deflection angle as a function of the impact parameter.}
    \label{fig:alphau}
\end{figure}

\begin{figure}[htpb]
    \centering
   \includegraphics[width=10cm, height=8cm]{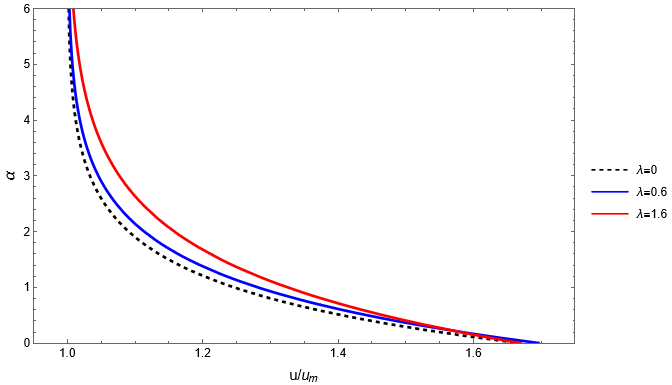}
    \caption{Graphical representation, for the Schwarzschild solution with holonomy corrections, of the deflection angle as a function of the reduced impact parameter.}
    \label{fig:alphaum}
\end{figure}

For the  Schwarzschild solution with holonomy corrections we consider the the following metric
\begin{equation}
    \begin{aligned}
    & A(x) = 1- \frac{1}{x}, \\
    & B(x) = \frac{1}{\left(1-\frac{1}{x}\right) \left(1-\frac{\lambda ^2}{\left(\lambda ^2+1\right) x}\right)}, \\
    & C(x) = x^2. 
    \end{aligned}
    \label{eq:SCholomy}
\end{equation}
\begin{equation}
    x_m = \frac{3}{2}
\end{equation}
\begin{equation}
    u_m = \frac{3 \sqrt{3}}{2}
\end{equation}

\begin{equation}
    \begin{aligned}
       & R(\zeta,x_0) = 2\sqrt{\frac{\lambda ^2+1}{\frac{\lambda ^2}{3}+1}}, \\
       &f(\zeta, x_0) = \frac{1}{\sqrt{\beta_1 \zeta +\beta_2 \zeta^2- \frac{\zeta^3}{x_0}}},
       \label{eq:fersch}
    \end{aligned}
\end{equation}
with the same $\beta_{1,2}$ from Eq. \eqref{eq:beta12}. This leads to
\begin{equation}
    b_1 =  \frac{R(0,x_m)}{2\sqrt{\beta_{2m}}} = \sqrt{\frac{\lambda ^2+1}{\frac{\lambda ^2}{3}+1}}.
    \label{eq:b1h}
\end{equation}
To compute $b_2$ we first write the function
\begin{equation}
    g(\zeta, x_m) = \frac{2}{z} \left(\frac{3 z \sqrt{\frac{\lambda ^2+1}{\lambda ^2+2 \lambda ^2 z+3}}}{\sqrt{z^2 (3-2z)}}-\sqrt{3} \sqrt{\frac{\lambda ^2+1}{\lambda ^2+3}}\right),
\end{equation}
which if integrated results in
\begin{equation}
    \int g(\zeta, x_m) d\zeta  = -2 \sqrt{3} \left(\sqrt{\frac{\lambda ^2+1}{\lambda ^2+3}} \ln (z)+\frac{2 \sqrt{\frac{\lambda
   ^2+1}{\lambda ^2+2 \lambda ^2 z+3}} \sqrt{\lambda ^2 (2 z+1)+3} \tan ^{-1}\left(\frac{\sqrt{\lambda
   ^2+3} \sqrt{3-2 z}}{\sqrt{\lambda ^2 (6 z+3)+9}}\right)}{\sqrt{\lambda ^2+3}}\right) + C.
   \label{eq:gholonomy}
\end{equation}
It is easy to see that the Schwarzschild result is immediately recovered if we use $\lambda=0$. The presence of the constant $\lambda$ in the above equation makes it impossible to determine the limit in $z=0$ analytically. Because of this limitation, we solve this integral numerically with values of the constant from zero to two. From this we can calculate the coefficient $b_2$, and show the result of the coefficients of the expansion of the deflection angle in Fig. \ref{fig:coef}.
In the Figs. \ref{fig:alphalambda}, \ref{fig:alphau} and \ref{fig:alphaum} we show respectively the deflection angle as a function of the constant $\lambda$, the impact parameter $u$ and the reduced impact parameter $u/u_m$.

\section{Observables for the Sagittarius A* \label{sec:obser}}

In this section we analyze the influence of the $\lambda$ parameter on the observables for both the weak and strong field regimes. We consider the mass and distance with respect to the black hole at the center of our galaxy, Sagittarius A*, which (we ignore uncertainties) is \cite{Richstone:1998ky,EventHorizonTelescope:2022wkp,GRAVITY:2023avo} 
\begin{equation}
\begin{aligned}
   & m = 4.297 \times 10^6 M_{\odot}, \\
    &D_{OL} = 2.55402 \times 10^{20} \,
    \end{aligned}
    \label{eq:valoresmedm}
\end{equation}

where $M_{\odot} =1.98892 \times 10^{30} \text{kg}$ is the Solar mass. Recall that we use geometric units unless otherwise noted. As suggested in Section \ref{sec:one}, the observables are the positions of the lensed images, the magnification, and the delay time between them. But before we turn to the numerical example, we need to clarify two points. First, we will work with flux rather than magnification (following  \cite{Bozza:2002zj,Keeton:2006sa}). In addition, we will consider different situations for each regime. For the weak field, we consider two lensed images, the first with positive parity and the second with negative parity. In the strong field, the photon can make  loops around the black hole before emerge into infinity and reach the observer. Therefore, we will consider observables that relate the first image and the contribution of all others.

\subsection{Weak Field}

\begin{figure}[htpb]
    \centering
   \includegraphics[width=10cm, height=8cm]{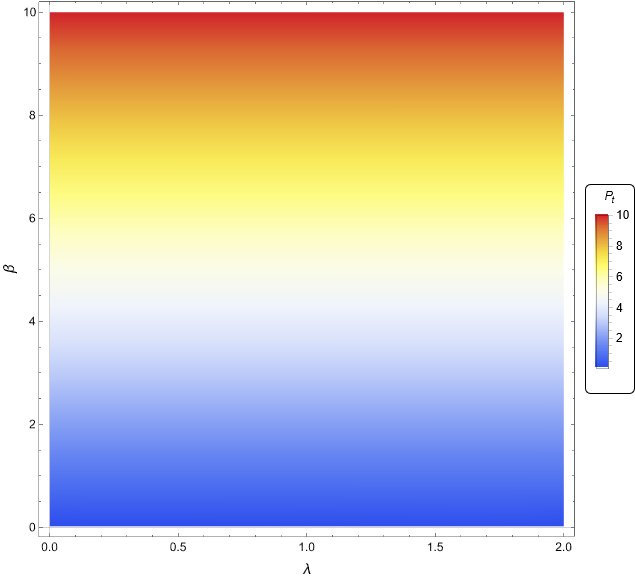}
    \caption{Density plot, for the Schwarzschild solution with holonomy corrections, of the angular distance between two lensed images $P_t$ ($m$ arc s ), given by the equation \eqref{eq:pt}, as a function of the angle $\beta$ and the parameter $\lambda$.}
    \label{fig:thetamais}
\end{figure}

\begin{figure}[htpb]
    \centering
   \includegraphics[width=10cm, height=8cm]{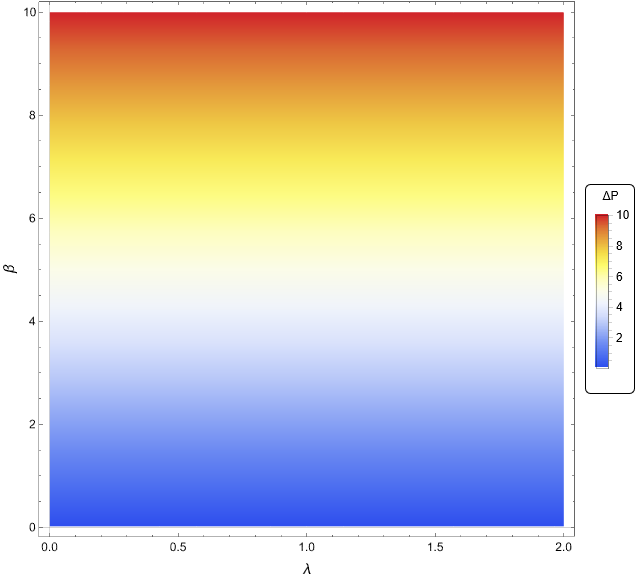}
    \caption{Density plot, for the Schwarzschild solution with holonomy corrections, of the difference of the angular positions between two  lensed images $\Delta P$ ($m$ arc s), given by the equation \eqref{eq:deltap}, as function of the angle $\beta$ and the parameter $\lambda$.}
    \label{fig:thetamenos}
\end{figure}

\begin{figure}[htpb]
    \centering
   \includegraphics[width=10cm, height=8cm]{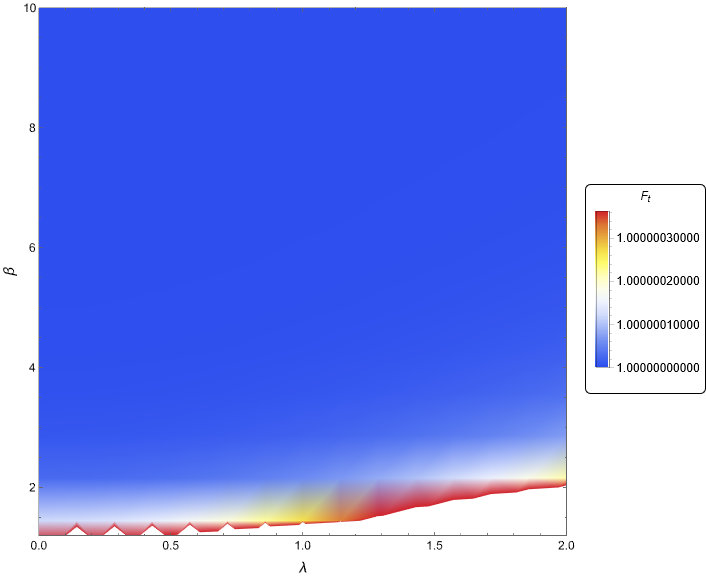}
    \caption{Density plot, for the Schwarzschild solution with holonomy corrections, of the total flux between two  lensed images $F_t$, given by the equation \eqref{eq:fluxt}, as function of the angle $\beta$ and the parameter $\lambda$.}
    \label{fig:fluxt}
\end{figure}

\begin{figure}[htpb]
    \centering
   \includegraphics[width=10cm, height=8cm]{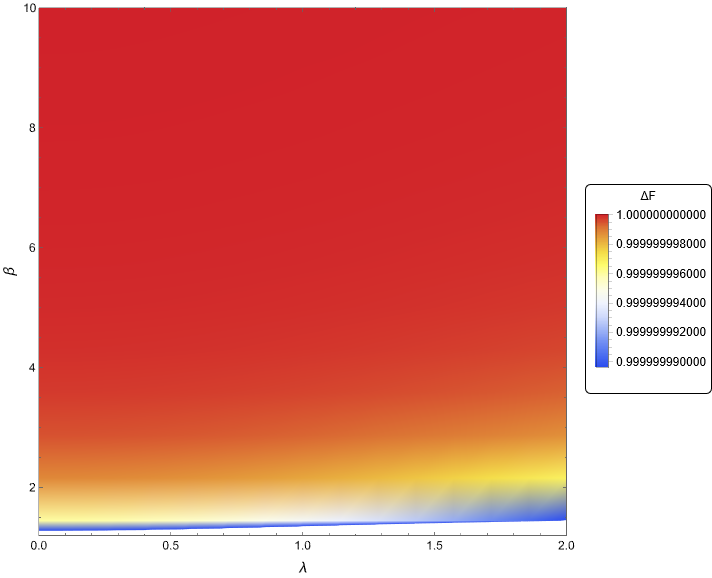}
    \caption{Density plot, for the Schwarzschild solution with holonomy corrections, of the difference of flux between two  lensed images  $\Delta F$, given by the equation \eqref{eq:deltaf}, as function of the angle $\beta$ and the parameter $\lambda$.}
    \label{fig:deltaflux}
\end{figure}

\begin{figure}[htpb]
    \centering
   \includegraphics[width=10cm, height=8cm]{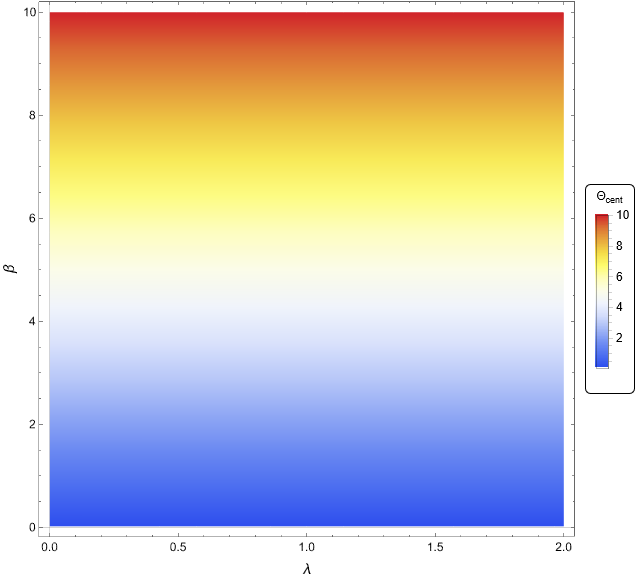}
    \caption{Density plot, for the Schwarzschild solution with holonomy corrections, of the differential time delay  between two  lensed images $\Theta_{cent}$ ($m$ arc s), given by the equation \eqref{eq:thetacent}, as function of the angle $\beta$ and the parameter $\lambda$.}
    \label{fig:centroi2}
\end{figure}

\begin{figure}[htpb]
    \centering
   \includegraphics[width=10cm, height=8cm]{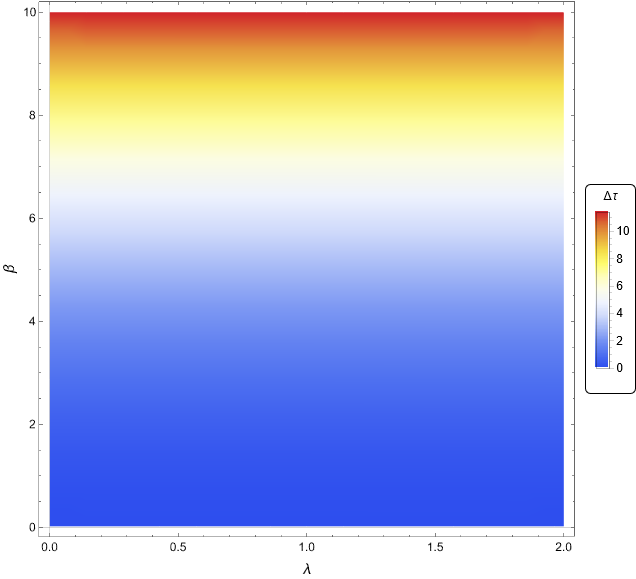}
    \caption{Density plot, for the Schwarzschild solution with holonomy corrections, of $\Delta \tau$ (s), given by the equation \eqref{eq:deltatau}, as function of the angle $\beta$ and the parameter $\lambda$.}
    \label{fig:delaywfbeta}
\end{figure}

For this regime we will consider a combination of the observables we worked on in section \ref{sec:one}, they are

\begin{equation}
    P_t = \vartheta^{+}+ \vartheta^{-} = \sqrt{A_1 \theta _E{}^2+\beta ^2}+\frac{\epsilon  A_2 \theta _E}{A_1},
    \label{eq:pt}
\end{equation}
\begin{equation}
    \Delta P = \vartheta^{+}- \vartheta^{-} = |\beta | -\frac{|\beta|    \theta _E}{\sqrt{A_1 \theta _E{}^2+\beta ^2}}\epsilon,
    \label{eq:deltap}
\end{equation}
\begin{equation}
    F_t = F^{+} + F^{- } = \frac{F_{src} \left(A_1 \theta _E{}^2+2 \beta ^2\right)}{2 |\beta|  \sqrt{A_1 \theta _E{}^2+\beta ^2}},
    \label{eq:fluxt}
\end{equation}

\begin{equation}
     \Delta F = F^{+}- F^{-} =  F_{src}-\frac{A_2 F_{src} \theta _E{}^3}{2 \left(A_1 \theta _E{}^2+\beta ^2\right){}^{3/2}}\epsilon
     \label{eq:deltaf}
\end{equation}

\begin{equation}
    \Theta_{cent} = \frac{\vartheta^{+}F^{+} - \vartheta^{-}F^{-} }{F_t} = \frac{|\beta|  \left(3 A_1 \theta _E{}^2+4 \beta ^2\right)}{2 A_1 \theta _E{}^2+4 \beta ^2}
    \label{eq:thetacent}
\end{equation}
\begin{equation}
    \Delta \tau = \frac{D_{OL} D_{OS} \left(\frac{1}{2} |\beta|  \sqrt{A_1 \theta _E{}^2+\beta ^2}+\frac{1}{4} A_1 \theta _E{}^2 \ln
   \left(\frac{\sqrt{A_1 \theta _E{}^2+\beta ^2}+|\beta| }{\sqrt{A_1 \theta _E{}^2+\beta ^2}-|\beta| }\right)+\frac{|\beta |
     A_2 \theta _E}{A_1} \epsilon\right)}{c D_{LS}}.
     \label{eq:deltatau}
\end{equation}
The flux is related to magnification by $F_i= |\mu_i| F_{src} $, where $F_{src}$ is the source flux. These equations  . To calculate this numerically, we use the equations \eqref{eq:thetaE} and \eqref{eq:novosangulos}. First, however, we need to convert the values in \eqref{eq:valoresmedm} to more appropriate units, i.e., parsecs (pc). We will use the following values
\begin{equation}
\begin{aligned}
   & m = 4.297 \times 10^6 M_{\odot} = 2.057 \times 10^{-7} \text{pc}, \\
    &D_{OL} = 2.55402 \times 10^{20} = 8.277 \text{kpc} \,
    \end{aligned}
    \label{eq:valoresmedmpc}
\end{equation}
this choice of units leads to $\vartheta_E = 0.0225813 \sqrt{d_{ls}}$ and $\epsilon =0.000181799/\sqrt{d_{ls}}$. we assume that $d_{ls}= D_{ LS }/ 1 (pc)$ to simplify notation. This distance is usually much smaller than the distance from the observer to the source $D_{ OS }$ and from the observer to the lens $D_{ OL }$. In other words,
$D_{ OL } \sim D_{ OS } >>D_{ LS }$. The following figures show the behavior of the practical observables as a function of the source angle $\beta$ and the loop quantum gravity parameter $\lambda$, we have $d_{ls} =1$ in all cases. We can see that the influence of $\lambda$ on the values of the observables in the considered intervals is practically zero. Figure \ref{fig:thetamais} shows the angular distance $P_t$ and Figure \ref{fig:thetamenos} shows the difference of the angular positions $\Delta P$, both increase with $\beta$. The total flux $F_t$,
shown in figure \ref{fig:fluxt}, and the flux difference $\Delta F$, shown in figure \ref{fig:deltaflux}, are constant for $\beta 2$; below this value we see that $F_t$ increases while $\Delta F$ decreases. However, these fluctuations are small.
 Figure \ref{fig:centroi2} shows the centroid $\Theta_{cent}$ and figure \ref{fig:delaywfbeta} shows the differential time delay $\Delta \tau$, both observables also increase with angle $\beta$.

\subsection{Strong Field}

 Here we focus on the asymptotic position approached by a set of images $\vartheta_{\infty}$, the distance between the first image (labeled $\vartheta_1$) and the others $s$,  the ratio between the flux of the first image and the flux of all the other images $r_m$ \cite{Bozza:2002zj}, and in the time delay between one photon with 2 loops from on photon with one loop around the lens \cite{Bozza:2003cp}. These are given by
\begin{equation}
    \vartheta_{\infty} = \frac{u_m}{ D_{OL}},
\end{equation}
\begin{equation}
    s = \vartheta_1 - \vartheta_{\infty}  = \vartheta_{\infty} e^{\frac{b_2-2\pi}{b_1} },
\end{equation}
\begin{equation}
    r_m = e^{\frac{2\pi}{b_1}},
\end{equation}
\begin{equation}
    \Delta T_{2,1} = [2 \pi -2\gamma] u_m + 2 \sqrt{\frac{B_m}{A_m}} \sqrt{\frac{u_m}{c_1}}e^{\frac{b_2}{2b_1}} \left(  e^{- \frac{ \pi \gamma}{b_1}}-  e^{- \frac{2 \pi \gamma}{b_1}}\right).
\end{equation}
In the time expression, $\gamma$ stands for the angular distance between the source and the optical axis as seen from the lens. In real observations, this angle should be of the order $\gamma \sim D^{-1}_{ OL }$. 
With the values in \eqref{eq:valoresmedm}, we obtain the data listed in Table \ref{tab:tabela1}. Since $\vartheta_{\infty}$ does not depend on $\lambda$, the result is $26.5807 \mu$arcsecs (the same as in the Schwarzschild case) and we therefore do not include it in the table. We note that the distance between the first and the other images increases with $\lambda$, and so does the delay time between them. The decrease in $r_m$ means that the first image becomes less intense compared to the other images as the parameter $\lambda$ increases.  In addition, it is clear that the delay time for this regime is  more intense when compared to the weak field regime.

\begin{table}[htpb]
    \centering
    \begin{tabular}{c|c|ccccc}
  & $\lambda$  & $b_1$ & $b_2$ & s ($\mu$arcsecs) & $r_m$ & $\Delta T_{2,1}$ (hours)  \\
\hline
 \text{} & 0 & 1. & -0.40023 & 0.0332657 & 6.82188 & 0.19431 \\
\hline
 \text{} & 0.1 & 1.00332 & -0.399568 & 0.0340313 & 6.79933 & 0.194315 \\
 \text{} & 0.2 & 1.01307 & -0.397839 & 0.0363476 & 6.73385 & 0.19433 \\
 \text{} & 0.3 & 1.02871 & -0.395748 & 0.0402631 & 6.63147 & 0.194354 \\
 \text{} & 0.4 & 1.04941 & -0.394252 & 0.0458289 & 6.50067 & 0.194385 \\
 \text{} & 0.5 & 1.07417 & -0.394335 & 0.0530645 & 6.35083 & 0.194423 \\
  & 0.6 & 1.10195 & -0.396812 & 0.0619262 & 6.19076 & 0.194466 \\
  & 0.7 & 1.13173 & -0.402209 & 0.0722898 & 6.02786 & 0.194512 \\
  & 0.8 & 1.1626 & -0.410737 & 0.083952 & 5.86776 & 0.194559 \\
  & 0.9 & 1.19382 & -0.422324 & 0.0966477 & 5.71435 & 0.194607 \\
  & 1 & 1.22474 & -0.436684 & 0.110078 & 5.57004 & 0.194654 \\
  & 1.1 & 1.25492 & -0.453397 & 0.12394 & 5.43611 & 0.1947 \\
  & 1.2 & 1.284 & -0.471976 & 0.137951 & 5.313 & 0.194745 \\
  & 1.3 & 1.31175 & -0.491927 & 0.151865 & 5.2006 & 0.194787 \\
  & 1.4 & 1.33803 & -0.512787 & 0.16548 & 5.09845 & 0.194828 \\
  & 1.5 & 1.36277 & -0.534139 & 0.178644 & 5.00589 & 0.194866 \\
  & 1.6 & 1.38595 & -0.555633 & 0.191246 & 4.92216 & 0.194901 \\
  & 1.7 & 1.4076 & -0.576979 & 0.203215 & 4.84648 & 0.194934 \\
  & 1.8 & 1.42775 & -0.59795 & 0.214512 & 4.77807 & 0.194965 \\
  & 1.9 & 1.44647 & -0.618373 & 0.225122 & 4.71622 & 0.194994 \\
  & 2 & 1.46385 & -0.638122 & 0.235048 & 4.66023 & 0.195021 \\
  \hline
 \end{tabular}
    \caption{Estimates for the main observables and the strong field limit coefficients for the black hole in the center of our galaxy  considering the Schwarzschild geometry with holomy corrections.}
    \label{tab:tabela1}
\end{table}

\section{Massive particles surface \label{sec:six}}

The photon surface concept used in section \ref{sec:two} also plays an important role in the study of black hole shadows. We will not deal with shadows here, but we would like to introduce a generalization of this concept, namely the surface of massive particles. In \cite{Kobialko:2022uzj} the authors propose a generalization of the photon surface for the case of massive and charged particles. They define a surface which also has the main property of a photon sphere, i.e:
any world line originally tangent to the surface of a massive particle remains tangent to it. The main difference between these two definitions is that the photon sphere formalism considers null geodesics with a fixed impact parameter, while for massive particles the fixed parameter is the total energy. A full description of this approach and its main implications can be found in \cite{Kobialko:2022uzj}. Here we restrict ourselves to the calculation of the surface  for a neutral particle of mass $m_0$ in a spacetime described by the holonomy-corrected Schwarzschild solution.

We start considering the static  metric tensor 
\begin{equation}
    ds^2 = -A dt^2 + Bdr^2 + C d \phi^2 + D d \theta^2 \,
\end{equation}
where $A,B,C$ and $D$ are free functions of $r$ and $\theta$, but we choose a surface with $r = const.$ 
The main equation of this method (for neutral particles) is
\begin{equation}
    \varepsilon_{\pm} = \pm m_o  \sqrt{\frac{\kappa^2\chi_{\tau}}{K}}.
    \label{eq:energym0}
\end{equation}
Here $\varepsilon$ is the total energy. According to   \cite{Kobialko:2022uzj}, $d \varepsilon/dr = 0$ defines marginally stable orbits, such as the Innermost Stable Circular Orbit  (ISCO). And the  value of $r$ at which the energy diverges defines the photon surface. In order to calculate the total energy \eqref{eq:energym0}, we must first calculate the second fundamental form and its trace, which (for this particular static spacetime) is defined as follows
\begin{equation}
    \chi_{\mu \nu} = \frac{1}{2 \sqrt{B}} \left( - \partial_r A dt^2 +  \partial_r C d \phi + \partial_r Dd\theta \right),
\end{equation}
\begin{equation}
    \chi = \chi^{\mu}_{\ \nu} = \frac{\partial_r \ln(ADC)}{2 \sqrt{B}}.
\end{equation}

Also, we have that the killing vector is $\kappa^{\mu}\partial_{\mu} = \partial_t$ and then $\kappa^{\mu}\kappa_{\mu} = - A$. And finally, we have
\begin{eqnarray}
    K = 3\chi_{\tau} - 2 \chi,
\end{eqnarray}
where $\chi_{\tau}$ for this static space time is
\begin{equation}
    \chi_{\tau} = \frac{2 \sqrt{A}}{r}.
\end{equation}

If we consider the  Schwarzschild metric \eqref{eq:metricasc}, for example, we have
\begin{equation}
\chi_{\tau} = \frac{2}{r}\sqrt{1- \frac{2m}{r}}, \ \ \ \ \chi = \left( \frac{1}{\sqrt{1- \frac{2m}{r}}} \right) \frac{2r-3m}{r^2}, \ \ \ \ \kappa^2 = - 1+ \frac{2m}{r}, \ \ \ \
K = 2 \left(\frac{1}{\sqrt{1- \frac{2m}{r}}} \right)  \frac{3m-r}{r^2}.
\label{eq:107}
\end{equation} 

So, the total energy for this case is
\begin{equation}
    \varepsilon^2/m_0^2 = \frac{(r-2m)^2}{r(r-3m)}.
    \label{eq:energysc}
\end{equation}
It is easy to see that the above expression diverges for $r =3m$, so we get the well-known result for the radius of the photon sphere. If we equate $d \varepsilon/ dr =0$, we get $r= 6m$, which represents the ISCO.

If we now proceed to the holonomy-corrected solution, we consider the metric \eqref{eq:metricah} and then have
\begin{equation}
         \chi = \frac{(2 r-3 m) \sqrt{\frac{\lambda ^2 (r-2 m)+r}{\left(\lambda ^2+1\right) (r-2 m)}}}{r^2}, \ \  \ \  K =  \frac{2 \left((2 r-3 m) \sqrt{1-\frac{2 m}{\left(\lambda ^2+1\right) (2 m-r)}}-3 r \sqrt{1-\frac{2
   m}{r}}\right)}{r^2}.
\end{equation}
The quantities $\chi_{\tau}$ and $\kappa^2$ are the same as \eqref{eq:107}. With these values, the energy is
\begin{equation}
    \varepsilon^2/m_0^2 = -\frac{r \left(1-\frac{2 m}{r}\right)^{3/2}}{(2 r-3 m) \sqrt{1-\frac{2 m}{\left(\lambda ^2+1\right)
   (2 m-r)}}-3 r \sqrt{1-\frac{2 m}{r}}} .
\end{equation}
We can simplify this expression by making the following change to the variable $R= r/m$. This leads us to
\begin{equation}
     \varepsilon^2/m_0^2 = \frac{2-R}{R \left(\frac{\sqrt{\frac{1}{\lambda ^2+1}} (2 R-3)}{R-2}-3\right)}.
\end{equation}

\begin{figure}[htpb]
    \centering
   \includegraphics[width=10cm, height=8cm]{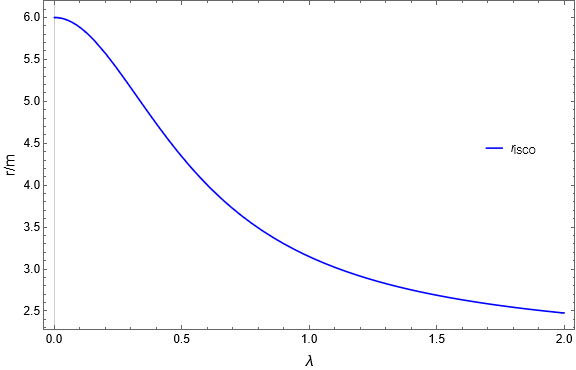}
    \caption{Radius of the massive particle surface as function of the holonomy parameter $\lambda$ given by the equation \eqref{eq:risco}}
    \label{fig:isco}
\end{figure}

Now, by taking the derivative of the above equation, we can find the radius of the massive particle surface. This operation results in
\begin{equation}
    r_{ISCO} = \frac{6m \left(\sqrt{\frac{1}{\lambda ^2+1}}-2\right)}{5 \sqrt{\frac{1}{\lambda ^2+1}}-6},
    \label{eq:risco}
\end{equation}
we plot this radius in the figure \ref{fig:isco}. Note that for $\lambda=0$ we get the result of Schwarzschild.  This decrease in $r_{ISCO}$ implies a disk of massive matter closer to the event horizon. Note that by defining the surfaces of photons and particles, we can construct the shadow and optical appearance of the black hole. However, we will address this issue in a later article.

\section{Summary and Discussion} \label{sec:conclusion}

In this article we describe the gravitational lensing effect for the Schwarzschild solution with holonomy corrections.
We treat the effect using the two types of limits commonly treated in the literature, i.e., the weak and the strong field regimes. Our main goal was to measure the influence of the parameter $\lambda$ associated with loop quantum gravity on the deflection angle and observables for each regime. We have used the method described by Keeton and Petters \cite{Keeton:2005jd}, where $\alpha$ is approximated by a power series around $u/m $. And for the strong field we use the formalism proposed by Bozza \cite{Bozza:2002zj}, where the deflection angle is given by a logarithmic approximation.

More specifically, in Sec. \ref{sec:one} we explored with the weak field regime, in which the formalism was summarized and applied it to the Schwarzschild solution, where the first term of the expansion, $\mathcal{A}_1 =4$, reduces to Einstein's result \cite{Weinberg}. Furthermore, we applied the formalism to the modified solution and found the expansion for the deflection angle in powers up to the order of $(m/u)^5$. An interesting detail is that the first order coefficient is different from four, i.e., $\mathcal{A}_1 \neq 4$, highlighting the fact that we are not working with classical general relativity. We note that the presence of the constant $\lambda$, a parameter related to loop quantum gravity (LQG), increases the deflection angle, as can be seen from the analysis of Fig. \ref{fig:alphawf}.
From the expression for $\alpha$, we derived the expression for the image position, magnification, total magnification, and position of the centroid (observables commonly discussed in the literature). We wrote the expression for these observables as an expansion in terms of the factor $\epsilon \theta_E/4D$, where $\theta_E$ is the Einstein angle. Keeton and Petters \cite{Keeton:2005jd} pointed out that in any hypothesis involving general relativity, the total magnification and the position of the center of gravity would not have first-order terms in this expansion, but here we can obtain these terms since $\mathcal{A}_1 \neq 4$.
We show the influence of the LQG parameter on the observables in the Figs. \ref{fig:theta012}--\ref{fig:delaywf}. 

Section \ref{sec:GB} was devoted to the proof that the coefficient $\mathcal{A}_1$ can be determined by a method other than the geodesic one. As described in \cite{Gibbons:2008rj}, the Gauss-Bonnet theorem reduces to the expression \eqref{eq:GWformula} in the case of gravitational lensing.
 We started from the same metric we used in the previous treatment and obtained the so-called optical metric and used it to calculate the Gaussian curvature $K$. As in the previous method, the integrals involved are not exactly solvable. We first used a Taylor series expansion that considers up to order $\lambda^2$ \eqref{eq:gbalpha} and later up to first order $m/u$ \eqref{eq:50}. This gave us exactly the same coefficient that we found earlier. It is interesting to note that this method leads to coefficients of higher order than the first, which are different from Eq. \eqref{eq:coefwf}. This is a point we would like to investigate in a future paper.

In Sec. \ref{sec:two}, we addressed the strong field limit.
As in the weak field regime, we began with a brief definition of the formalism used to obtain the approximation to this regime. We showed that the expression \eqref{eq:explog}, which holds for any static and spherically symmetric metric, is an approximation of the deflection angle when $x_0 \rightarrow x_m$ (where $x_0$ is the distance of closest approach and $x_m$ is the capture radius or the radius of the photon sphere), or in terms of the impact parameter $u \rightarrow u_m$. We applied the formalism to the Schwarzschild solution ($x_m = 1.5$ ) and showed in Fig. \ref{fig:comp} the exact deflection angle (obtained by numerical integration of \eqref{eq:alphageral}) and the approximations for weak and strong fields for this solution. We then applied the formalism to the Schwarzschild solution with holonomy corrections.  
The radius of the photon sphere is the same as the previous one since it does not depend on the metric function $B(r)$, therefore, $u_m$ will also be the same. The coefficient $b_1$ \eqref{eq:b1h} can be obtained easily, however, the integral \eqref{eq:gholonomy} cannot be evaluated to zero for any value of the parameter $\lambda$. Therefore, we computed the coefficient $b_2$ numerically using the interval $0 < \lambda <2$. Figure \ref{fig:coef} showed the variation of the coefficients as a function of the LQG parameter. In Figs. \ref{fig:alphalambda}--\ref{fig:alphaum} we showed the behavior of the deflection angle with respect to $\lambda$. We verified that, as in the weak-field regime, it increases with increasing parameter.

In Sec. \ref{sec:obser} we calculate the observables using a numerical example with experimental data of the black hole at the center of our galaxy (Sagittarius A*) \cite{Richstone:1998ky,EventHorizonTelescope:2022wkp,GRAVITY:2023avo}.
For the weak field regime, we focus on a situation with two images, one with positive parity and the other with negative parity. The observables in this case are the angular separation $P_t$, the difference in angular positions $\Delta P$, the total flux $F_t$, the flux difference $\Delta F$, the centroid $\Theta_{cent}$, and the differential time delay $\Delta \tau$. Figures \ref{fig:thetamais}, \ref{fig:thetamenos}, \ref{fig:fluxt}, \ref{fig:deltaflux}, \ref{fig:centroi2}, and \ref{fig:delaywfbeta} show a plot of these quantities as a function of the angle of the source $\beta$ and the LQG parameter. We see that the influence of $\lambda$ on these quantities is small. For the strong field regime, we focus on the asymptotic position approached by a set of images $\vartheta_{\infty}$, the distance between the first image (denoted $\vartheta_1$) and the other $s$, the ratio between the flux of the first image and the flux of all other images $r_m$, and on the time delay between a photon with 2 loops and a photon with one loop around the lens $\Delta T_{2,1}$. We find that the observable $\vartheta_{\infty}$ does not depend on $\lambda$, $s$ and $\Delta T_{2,1}$ increase with it, while $r_m$ decreases.  We showed in Table \ref{tab:tabela1} the numerical result obtained. There are already attempts to measure the strong gravitational lensing effect \cite{Nightingale:2023ini}. We therefore hope that in the near future it will be possible to use this formalism to select a particular model of a black hole via the  coefficients of the strong field.

In Sec. \ref{sec:six} we have considered the extension of the concept of photon surface described in \cite{Kobialko:2022uzj} and applied it to the corrected Schwarzschild solution.
We determined the radius of the innermost stable circular orbit (ISCO) and found that it decreases with increasing parameter $\lambda$.
This form of obtaining particle surfaces with the same properties as the photon sphere may be of interest in a future work where we plan to address the shadows of this solution.

\acknowledgments

MER  thanks Conselho Nacional de Desenvolvimento Cient\'ifico e Tecnol\'ogico - CNPq, Brazil, for partial financial support. This study was financed in part by the Coordena\c{c}\~{a}o de Aperfei\c{c}oamento de Pessoal de N\'{i}vel Superior - Brasil (CAPES) - Finance Code 001.
FSNL acknowledges support from the Funda\c{c}\~{a}o para a Ci\^{e}ncia e a Tecnologia (FCT) Scientific Employment Stimulus contract with reference CEECINST/00032/2018, and funding through the research grants UIDB/04434/2020, UIDP/04434/2020, CERN/FIS-PAR/0037/2019 and PTDC/FIS-AST/0054/2021.


\end{document}